\documentclass[a4paper,11pt]{article}
\usepackage{graphicx}
\usepackage[space]{grffile}
\usepackage{latexsym}
\usepackage{textcomp}
\usepackage{longtable}
\usepackage{tabulary}
\usepackage{booktabs,array,multirow}
\usepackage{amsfonts,amsmath,amssymb}
\usepackage{natbib}
\usepackage{url}
\usepackage{hyperref}
\hypersetup{colorlinks=false,pdfborder={0 0 0}}
\usepackage{etoolbox}
\makeatletter
\usepackage{authblk}
\usepackage[utf8]{inputenc}
\usepackage[english]{babel}
\usepackage[top=1in, bottom=1in, left=1in, right=1in]{geometry}
\title{Turbulence and equatorial waves in moist and dry shallow water systems}
\date{}
\author[1]{J. Schr\"ottle  \thanks{Corresponding author: J. Schr\"ottle, josefs@mail.tau.ac.il}}
\author[2]{D.L. Suhas}
\author[1]{N. Harnik}
\author[2,3]{J. Sukhatme}
\affil[1]{Department of Geophysics, School of Earth Sciences, Tel Aviv University, Israel.}
\affil[2]{Centre for Atmospheric and Oceanic Sciences, Indian Institute of Science, Bangalore, 560012, India.}
\affil[3]{Divecha Centre for Climate Change, Indian Institute of Science, Bangalore, 560012, India.}

\begin{document}

\maketitle

\begin{abstract}
Turbulence and large-scale waves in the tropical region are studied using the spherical shallow water equations. With mesoscale vorticity forcing, both moist and dry systems show kinetic energy scaling that is dominated by rotational modes, has a $-5/3$ exponent.
At larger planetary scales, the divergent component of the energy increases and we see a footprint of tropical waves. The dry system shows a signature of the entire family of equatorial waves, while the moist simulations only show low frequency Rossby, Kelvin and mixed Rossby gravity waves with an equivalent depth that matches rapid condensation estimates. Further, runs with interactive moisture exhibit spontaneous aggregation with the equilibrium moist energy spectrum obeying a $-2$ power-law. Synoptic scale moisture anomalies form in heterogeneous background saturation, and are sustained by advection and convergence, within rotational gyres that dominate the tropical region. 
In that case, along with equatorial waves and turbulence, a large-scale eastward moving moisture wave appears in the midlatitudes. In all simulations with upscale energy transfer, a systematic equatorial zonal mean zonal flow develops which is easterly and westerly for the dry and moist ensembles, respectively.
The interaction of this zonal mean flow with a spatially heterogeneous saturation field results in the formation of a moist stationary tropical wave. The super-rotating flow is driven by rotational eddy momentum fluxes due to enhanced equatorial Rossby wave activity in the moist runs. 
Interestingly, even though Kelvin waves are more energetic in the dry simulations, the easterly momentum flux is also due to rotational eddy fluxes. The nature of eddies is such that the tropical circulation in the dry and moist cases tends to homogenize and exaggerate potential vorticity gradients, respectively.  
These experiments demonstrate the co-existence of tropical waves and turbulence, and highlight the fact that the vortical and divergent wind are inextricably linked with the evolving moisture field.

\noindent \textit{Moist Turbulence, Moist Shallow Water Equations, Equatorial Waves, Aggregation, Equatorial Jets}

\end{abstract}

\section{Introduction}

\noindent Waves in the tropical atmosphere appear as coherent cloud patters and have been measured by geostationary satellites at temporal resolutions ranging from days to intraseasonal time scales \citep[e.g.,][]{takayabu,wheeler1999}. Specifically, waves have been identified in wavenumber-frequency analyses of outgoing longwave radiation (OLR) that is often used as a proxy for deep convection and even other variables such as zonal winds from reanalysis data \citep{Hendonwheeler2008}. Satellite data also allows for the detection of slowly propagating bands of water vapor in a relatively dry atmospheric column \cite{schroettle2020}. 
Some of the identified tropical waves correspond to classical dry modes or solutions of shallow water systems in the equatorial region \citep{matsuno1966} --- albeit, with a smaller equivalent depth \citep{Kiladis-rev}. While a theoretical estimate of the reduced equivalent depth, especially outside the so-called rapid condensation limit, is a challenge, lower wave speeds are understood to be due to the dynamical coupling with water vapor. Indeed, the importance of moisture, and its advection, for convectively coupled equatorial waves \citep[CCEWs;][]{WKW} has been highlighted in controlled cloud resolving simulations \citep{kuang2}. The generation of such slow CCEWs has been noted in full complexity and simplified general circulation models \citep[GCMs;][]{lin1,Frier}. Of course, mismatches between observations and full blown models in terms of wave speeds and their structure are the subject of ongoing scrutiny \citep{lin2}. There is also an extensive body of literature on simplified models, i.e., those with one or two vertical modes and prescribed heating, that explore linear instabilities to account for the generation of CCEWs \citep[see, for example,][]{mapes,maj-she}. Such linearized models with simplified vertical structure but explicit moisture evolution have also been examined in this context \citep{kuang1,KhMa1}. Interestingly, equatorial Rossby waves are not included in these efforts and are usually dealt with separately \citep{CG}. Though, both tropical Rossby and Kelvin waves with suitably reduced speeds have been observed in initial value problems with a nonlinear moist shallow water system \citep{suhas2020}.  It is important to note that these waves exist at large spatial scales, indeed, most wavenumber-frequency diagrams display variability from approximately wavenumber 1 to 10, or from four to five thousand kms to the planetary scale \citep{wheeler1999,WKW,Hendonwheeler2008}.

\noindent At relatively smaller scales, from about 50 to 2000 km, i.e., the mesoscales through to synoptic scales, the atmospheric kinetic energy (KE) is observed to follow power-law scaling \citep{NG1985}. Specifically, in the midlatitude upper troposphere KE follows $-5/3$ and $-3$ exponents in the range of 50--500 km (mesoscales) and above (up to about 2000 km; synoptic scales), respectively \citep{NG1985}. The steeper $-3$ exponent is generally accepted to be a result of a forward enstrophy transfer quasi-geostrophic regime \citep{Charney,BoerS,Bartello}. The mesoscale range is associated with a forward energy transfer \citep{ChoLind,LindCho}, though there are different candidates --- ranging from rotating stratified turbulence with wave mode dominance \citep{Kitamura,SS,Valgren}, purely stratified dynamics \citep{L7,LindBerth}, balanced surface quasi-geostrophic turbulence \citep{TS} to inertia-gravity waves \citep{Callies} --- proposed for the $-5/3$ scaling. The correct choice from among these possibilities depends on the ratio of energy in rotational and divergent components of the horizontal mesoscale flow field \citep{Callies,Lindborg2015}, which itself appears to be non-universal \citep{Bierdel}. 

\noindent In the tropics, the situation regarding the mesoscale spectrum is likely more complicated due to the effects of moisture on turbulence. In fact, in a $f-$ plane study, \citet{WaiteS} noted that injection of energy due to latent heating enhanced the role of divergent components in moist baroclinic waves. On the other hand, numerical simulations of tropical cyclones showed a forward energy transfer and $-5/3$ KE power-law dominated by rotational modes, at scales below 500 km in the upper troposphere \citep{wang2018}. But, {\it in situ} aircraft data taken from flights through hurricanes paint a more diverse picture with mesoscale slopes going from $-5/3$ to $-3$ depending on the strength of systems \citep{Vonich}. More broadly, simulations using the Weather Research and Forecast (WRF) model, initialized with temperature anomalies in a moist environment show that buoyancy via moist convection plays an important role in the mesoscale spectrum in the tropical region \citep{Sunetal}. Here, rotational and divergent modes were observed to have an almost equal contribution to the total horizontal KE of the flow, both of which scaled with a $-5/3$ exponent. But, clear cascades could not be identified and buoyancy was noted to inject energy at all scales. Interestingly, two-dimensional moist stratified turbulence also showed the establishment of $-5/3$ scaling from an initially flat spectrum via (unequal) energy growth at all scales \citep{sukhatme2012}.

\noindent Given this situation, we study turbulence (both moist and dry) using a shallow water system with an eye towards the mesoscales, synoptic and planetary scales in the tropical atmosphere. In essence, rather than isolating any one of these ranges, given their interactions, we hope to simulate them simultaneously. The shallow water system affords us the possibility of resolving this extended range of scales efficiently and is also possibly the simplest model to include the relevant dynamical ingredients required to make a connection to the tropical atmosphere. In a turbulent context, explorations of the decaying and forced-dissipative spherical shallow water systems have been without moisture and have almost exclusively focused on extratropical phenomena such as the emergence of jets and vortices \citep[see, for example,][]{scott-pol-swe}. Though, in recent years, moist shallow water systems have been quite widely used in a tropical context. For example, studies aimed at the Madden-Julian Oscillation and other intraseasonal modes have used the shallow water framework with condensation, evaporation and either explicit \citep{RoZe1,RoZe2,Vallis}, or implicit 
\citep{Solod,Yang} treatment of water vapour. In addition,  the roles of moisture gradients \citep{Sobel,Sukhatme2,joyQG,suhas2020} and convergence in the boundary layer \citep{wang2016} have also been studied. All of these studies focus on the generation of large-scale waves and do not consider turbulent solutions. Here, the specific questions we address include, (i) Can large-scale equatorial waves be excited with small-scale forcing through an upscale transfer of rotational KE?
(ii) What are the differences in the energy transfer and excited large-scale waves (if any) between dry and moist turbulence? (iii) How does the rotational and divergent make up of the turbulent solution differ when moisture enters the dynamics? (iv) Does self-aggregation of moisture take place spontaneously? (v) How does the energy transfer among different scales depend on the nature of the forcing employed? For example, given both of their physical relevance, how does vorticity forcing differ from divergence forcing? In all, the answers to these questions should help in developing an appreciation for the establishment of a hierarchy of structures from mesoscales to planetary scales, and the role of dynamically interactive moisture in the establishment of these tropical atmospheric features. Further, the scaling of spectra that emerge here could be of use in interpreting analogous results from more complicated models or observations of the tropical atmosphere.

\noindent The following section in this paper introduces the stochastically forced shallow water equations and the setup of our numerical experiments. Results from small scale dry vorticity and divergence forced simulations are presented in Section~3, then Section~4 focuses on equatorial waves, energy cascades, and self-aggregation in homogeneous and heterogeneous moist backgrounds. Section~5 describes the evolution of mean flows, before we close with a brief summary and an outlook in Section~6.

\section{Overview of shallow water simulation setup}

We consider the moist shallow water equations on a sphere with isotropic stochastic forcing 
for a single layer in the lower atmosphere on top of a moist saturated aqua planet. In vorticity divergence form these equations read \citep{bouchut2009}, 
\begin{align}
\partial_t \zeta = -\nabla \cdot (\zeta  {\bf v}) + f_\zeta - \zeta/\tau_\zeta \nonumber, ~~~~~~~~~~~~~~~~~~~~~~~~~~~~~~~~~~~~~~~~~~~~~~~~\\
%
\partial_t \delta = \nabla \times (\zeta {\bf v}) \cdot \hat{k}  - \Delta \left(gh + {\bf v}^2 /2\right) + f_\delta - \delta/\tau_\delta \nonumber,~~~~~~~~~~~~~~\\
%
\partial_t h = -\nabla \cdot (h {\bf v}) - L\,q^{+} / \tau_c  - L\,q^{-} / \tau_e + f_h - (h-H)/\tau_h \nonumber,\\
%
\partial_t q = -\nabla \cdot (q {\bf v}) + f_q -q^{+} / \tau_c -q^{-} / \tau_e.~~~~~~~~~~~~~~~~~~~~~~~~~~~~~ 
\label{a1}
\end{align}
\noindent Here, $\zeta$ is the absolute vorticity and $\delta$ is the divergence of the horizontal wind $\bf v$. In the divergence equation, $\nabla \times (\zeta {\bf v})  \cdot \hat{k}$ is the vertical component of the curl. The height of the layer is $h$ and $H$ is its global mean. The moisture $q$ is composed of a background state ($q_s$) and a perturbation ($q'$). The geopotential is $gh$ and local kinetic energy is ${\bf v}^2 /2$, where $g$ is the gravitational constant of 9.81\,m\,s$^{-2}$. $\Delta$ is the two-dimensional Laplacian. Linear drag in vorticity, divergence, and height equation mimic friction and radiative damping with a time scale $\tau_\delta = \tau_\zeta = \tau_h$ of 500 days. These terms prevent the pile up of energy by removing energy from large scales. 
In the passive tracer limit, latent heat release $L=0$, condensation and evaporation time scale $\tau_c, \tau_e \rightarrow \infty$. The moisture equation has a source and sink in the form of evaporation and condensation. These follow a Betts-Miller type protocol whose specific forms of $(q^+,q^-)$ are discussed below. The forcing components are 
$f_\zeta$ in vorticity and $f_\delta$ in the divergence equation. Thermodynamic forcings on height and moisture fields are represented with $f_h$ and $f_q$, respectively.
In the absence of forcing, drag and dissipation the system conserves moist potential vorticity \citep{joyQG}. In fact, when the background saturation field is allowed to vary in space, the conservation of moist potential vorticity proves to be a useful guideline for the generation and propagation of moist Rossby waves \citep{suhas2020}.

\noindent All experiments with a dry atmosphere are run with a global mean depth of 100\,m, this keeps the shallow water wave speed at least one order of magnitude below the speed of sound waves. To allow for moist Kelvin waves with typical phase speeds close to tropical values of approximately 20\,m s$^{-1}$, we choose $Q=\max(q_s)=50$\,g kg$^{-1}$ and $L\leq1$\,m / g kg$^{-1}$, with a resulting reduced shallow water depth of 50\,m for all moist experiments. To remove energy from the flow at small scales, we apply hyperdiffusion as in classic approaches of stochastically forced turbulenct flows \citep{mcwilliams1984,frisch1984,vallis1993}. The equations are solved on a sphere by employing the efficient spectral transform library SHTNS \citep{schaeffer2013}. The equations are solved on a grid consisting of $512 \times 256$ mesh points in longitude and latitude, while using an advection time step of $\Delta t = 50$\,s. Stochastic forcing is adopted from simulations of the dry barotropic vorticity equation \citep{suhas2015}. Note that the library and the stochastic forcing have been employed before to study dry equatorial superrotation and moist initial value problems in shallow water system \citep{suhas2017,suhas2020}.

\subsection{Key features of the forcing fields in the moist and dry models}

When the forcing is uncorrelated in time, it can be described as a L\'{e}vy process, or a Markov process of lowest order. The forcing can turn into a Markov process of higher order, when random numbers between time steps are explicitly correlated. The forcing is formulated with mathematical rigor in Appendix~A and always exhibits a random phase. 
Spectral noise is conservative and formulated as Laplacian of a time-dependent stream function $\Psi$ to represent vorticity forcing or potential $\Phi$ to represent divergence forcing, that continuously forces at a defined wave number range around $k_0\pm2$.
By construction, the direct forcing of vorticity would lead to an immediate wind field free of divergence and vice versa.
Furthermore, super-positions of the forcing exhibit the same properties, as the forcing is linear. The only way to create divergence or vorticity that was not forced directly is by non-linear effects as advection or interaction with planetary vorticity.
In this work, we are interested in the formation of large scales from small scale forcing, i.e., forcing is typically in the mesoscales at approximately 400 km. Further, we mainly force the vorticity and divergence fields as these represent physically meaningful ways of injecting energy into the system. Specifically, forcing via the vorticity equation is a means of representing the effect of smaller-scale unresolved dynamics on the column integrated shallow water system \citep{scott2008equatorial}. On the other hand, divergent forcing, which represents the influence of convective events, is a way to mimic the small scale nature of the real-world divergent field \citep{koshyk2001horizontal}. As is known, the adjustment of the shallow water system can be quite different depending on the type of initial imbalance \citep{kuo-polvani}, thus, in addition to vorticity and divergence forcing, we also consider other types of forcing functions (i.e., $f_h$ and $f_q$) to develop a more complete feel for the system at hand. 

\noindent Moisture is allowed to fluctuate around the saturated state $q_s$. Extending the parameterization by \citet{gill1982}, we also allow for evaporation.
Latent heat is released, when the moist column is oversaturated: $q^{+} := q -  q_s$, where $q - q_s > 0$. We assume all the water stays in the column as cloud droplets, allowing for re-evaporation and cooling 
when the column is undersaturated: $q^{-} := q - q_s$, where $q - q_s< 0$. The
saturation field ($q_s$) can be isotropic and constant throughout the simulations \citep{bouchut2009,RoZe1,RoZe2,RoZe3}, 
or be a prescribed function of space \citep{Sobel,Sukhatme2,joyQG,suhas2020}.
Condensation occurs within a time scale $\tau_c$ and evaporation at a time scale $\tau_e$. Observations suggest a time scale of $1-12$\,h for $\tau_c$ \citep{frierson2004}.
We also ran the code assuming all condensed droplets fall out as rain, and evaporation from the ocean balances precipitation \cite{gill1982}, represented in our model by removing the term $-Lq^-/\tau_e$. However, these simulations don't reach a stable equilibrium when we applied a stochastic continuous mesoscale forcing on the moisture or vorticity field, which is probably due to the inability of the slow radiative damping to compensate for the condensational mass loss. We note that the mass loss could be balanced in such a scenario by making the saturation $q_s$ depth and consequently time dependent as in previous work \cite{frierson2004}. This dependency would counter balance the moist coupling of the thermodynamic equation similarly to a damped harmonic oscillator due to its restoring effect in the following way: the gradient would decrease the flux of mass of the layer through adjusting condensation when $h$ decreases, as well as increase the flux of mass into the layer through adjusting condensation when $h$ decreases. 
Moist shallow water experiments by \citet{suhas2020} with this numerical code have shown that results are independent of choosing the evaporation and condensation time scale, as long as both are smaller than one day. We set the evaporation time scale for most of the presented experiments to $\tau_e = 1$\,day and $\tau_c = 1/5\,$day, taking into account that evaporation of cloud droplets takes more time than the formation of clouds through condensation in deep convection. Though, in some cases, we have also used $\tau_c=\tau_e$ and variations of these timescales from a few hours to a day do not affect the results.  All simulations represent a lower layer in the atmosphere, so a sink term in the equation for its depth can also be thought of as a mass loss term due to convective updrafts, or a gain term may occur due to downdrafts.

\noindent Overall, simulations take about a few hundred days to reach a quasi-equilibrium state. Here, all simulations are run up to at least 1000 days to create appropriate fields for statistical post-processing. Mean profiles and spectra are presented for single days (to show the development of turbulence), as well as averaged over several days within a time period of a few hundred days during quasi-stable equilibrium. For a more robust picture, we run 80 ensemble simulations with random isotropic forcing of the same magnitude. A detailed description of all experiments performed is presented in Table \ref{table:experiments}. 

\section{Dry Simulations: Turbulence and Waves}

\begin{figure}
  \centering
    \includegraphics[width=1.\textwidth]{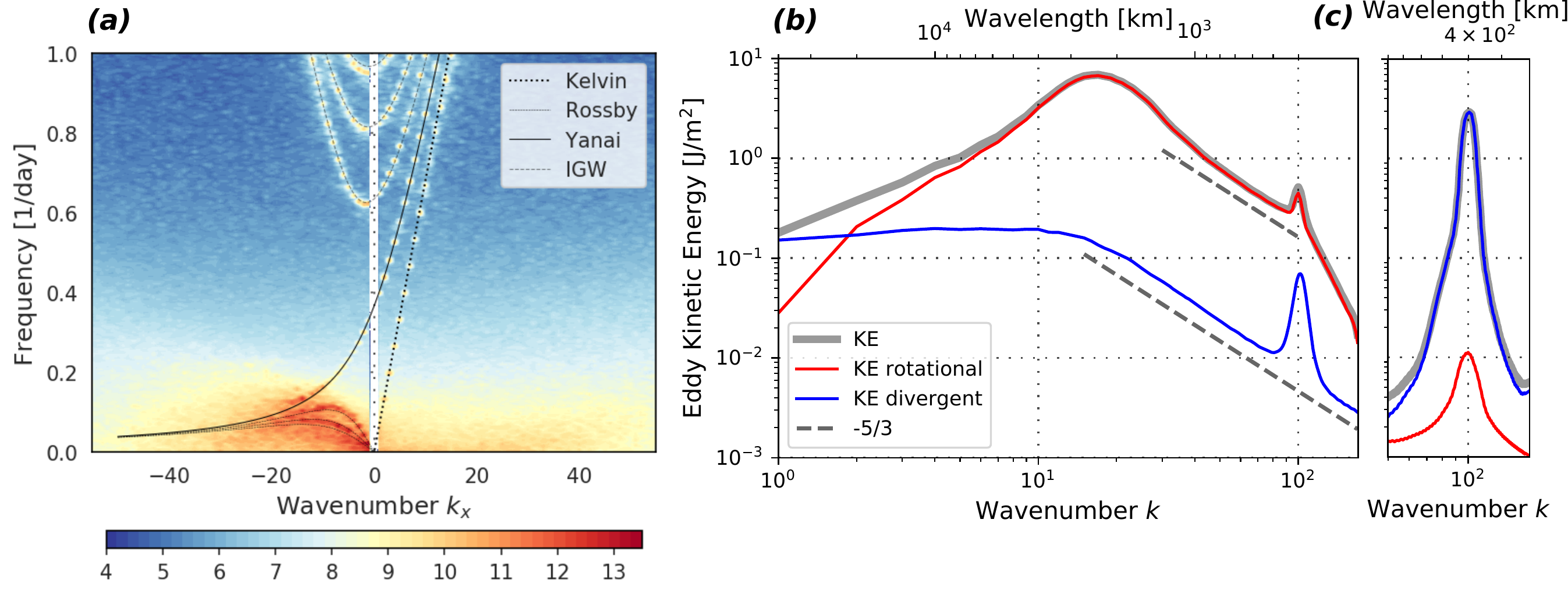}
  \caption{Dry small-scale vorticity forcing. {\bf (a)} Wavenumber-frequency diagram showing eddy PE from day 500 to 800 with a sampling frequency of 4\,h$^{-1}$ within an equatorial region between $\pm15^\circ$\,North/South. Solid lines follow the theoretical dispersion relations of Kevlin, Rossby, Yanai, and Inertia Gravity Waves. {\bf (b)} Spatial spectra of vortical, divergent \& total eddy KE for the tropical region $\pm20^\circ$\,North/South are plotted averaged from day 200--1000 with daily sampling. Two lines with $-5/3$ slopes are drawn for reference. {\bf (c)} KE spectra for the dry small-scale divergence forced experiment.}
  \label{fig:VrtFrSmallScaleDryKeSpectra}
\end{figure}

\noindent We begin with the dry equations and isotropic vorticity forcing at small scales (wave number 100, i.e., approximately 400 km) over the entire sphere. The stationary tropical wavenumber-frequency spectrum from this simulation is shown in Figure \ref{fig:VrtFrSmallScaleDryKeSpectra}a. 
Near the classical dispersion relations for Rossby, Yanai, Kelvin, and inertia gravity waves \citep{matsuno1966}, the wavenumber-frequency diagram shows increased intensity at discrete points in space and time \citep{vallis2017,ofer2020,suhas2020}. 
Interestingly, the waves appear as clear maxima aligned with theoretical linear dispersion relations on top of a red background spectrum. Further, almost all the wave activity is concentrated at large scales, i.e., below wavenumber 20, or, above $\approx$3500 km. The signature of waves at such large scales is consistent with idealized, three-dimensional, triply periodic $f-$ plane numerical experiments \citep{Asselin}. We emphasize that no background has been removed in Figure \ref{fig:VrtFrSmallScaleDryKeSpectra}a, indeed the high sampling rate allows the individual peaks to stand out over the background.

\noindent On a $f$- plane, vortical triads of the shallow water system form the quasi-geostrophic equations and support an inverse transfer of KE \citep{rem-smith}, as has been confirmed in  numerical simulations \citep{Farge,Yuan}. 
As seen in Figure \ref{fig:VrtFrSmallScaleDryKeSpectra}b, in the equatorial region, the continuous forcing of vorticity with white noise in time at small scales also initiates an upscale transfer of energy that persists till the equatorial deformation scale ($\sqrt{c/\beta}$; approximately wavenumber 20 or 3500 km, where $c$ represents the shallow water wave speed). We emphasize that this represents the distribution of energy among different scales in the tropics as we have set the velocity field to zero outside $\pm 20^\circ$ before computing the spectrum. Here, KE is primarily composed of vortical modes and follows a $-5/3$ slope. In the tropics the appropriate suite of interactions, say for example on an equatorial $\beta$- plane, includes Rossby triads \citep{ripa}, and these appear to result in the inverse transfer in the vortical component seen in Figure \ref{fig:VrtFrSmallScaleDryKeSpectra}b.
In particular, much like incompressible two-dimensional turbulence \citep{kraichnan1967}, under small-scale forcing this $-5/3$ scaling represents an inverse energy transfer regime of geostrophic turbulence \citep{Charney}. 
We note that similarly forced quasi-linear versions of this run did not lead to an efficient transfer of energy to larger scales, instead they led to an accumulation of vortical eddy kinetic energy at the forcing scale (Table~\ref{table:experiments}). This supports the conclusion that the $-5/3$ scaling is a result of non-linear eddy-eddy interactions.
Interestingly, though the divergent component has a comparatively much smaller amount of energy up to the deformation scale, we observe that it too scales with a $-5/3$ exponent and shows a pile up of energy at the forcing scale. The energy in these two components is consistent with estimates of the ratio of divergent to rotational energy in equatorial Rossby waves \citep{delayen-yano}.

\noindent As mentioned, the self-similar scaling of KE persists from the forcing scale to about wavenumber 20 (or a length scale of approximately 3500 km). Above this, vortical eddy kinetic energy begins to decrease monotonically. At these small wave numbers (large scales), the divergent modes gain strength and this coincides with the wave number range where equatorial waves were identified.
Overall, the vortical energy dominates over the divergent contribution in the tropics \citep{Yano-asy}, but at planetary scales, we find that the excited divergent modes are energetically comparable to the vortical modes. This is in line with the notion that particular classes of waves in the tropical atmosphere have a prominent divergent contribution at large scales \citep{Yasu-mapes}. Of course, it should be kept in mind that these largest scales are likely to be sensitive to the damping employed in the numerical simulation. 

\noindent For comparison, the KE spectrum for the run with small-scale divergence forcing is shown in Figure \ref{fig:VrtFrSmallScaleDryKeSpectra}c. In contrast to vorticity forcing (Figure \ref{fig:VrtFrSmallScaleDryKeSpectra}b), energy remains trapped at the forcing scale. There are no signs of interscale energy transfer and energy is largely confined to the divergent component suggesting an ineffective transfer from the directly forced divergent flow to rotational modes. The response to height forcing in the dry system is similar to divergence forcing (Table \ref{table:experiments}). 
At first glance, this seems to be in contrast to \citet{ofer2020}, who used temporally correlated stochastic forcing of layer depth in a dry shallow water model and excited equatorial waves. But, they also did not observe inverse transfer with a characteristic slope of $-5/3$. In fact, the maximum energy in their runs also remained trapped around the forcing scale as in our divergence and height forcing scenarios. 
At equilibrium both forcing types lead to global mean eddy PE and divergent KE of equal magnitudes for the white noise (in time) forcing (Table~\ref{table:experiments}), as would be expected from small-scale linear gravity waves.

\begin{figure}
  \centering
    \includegraphics[width=1.\textwidth]{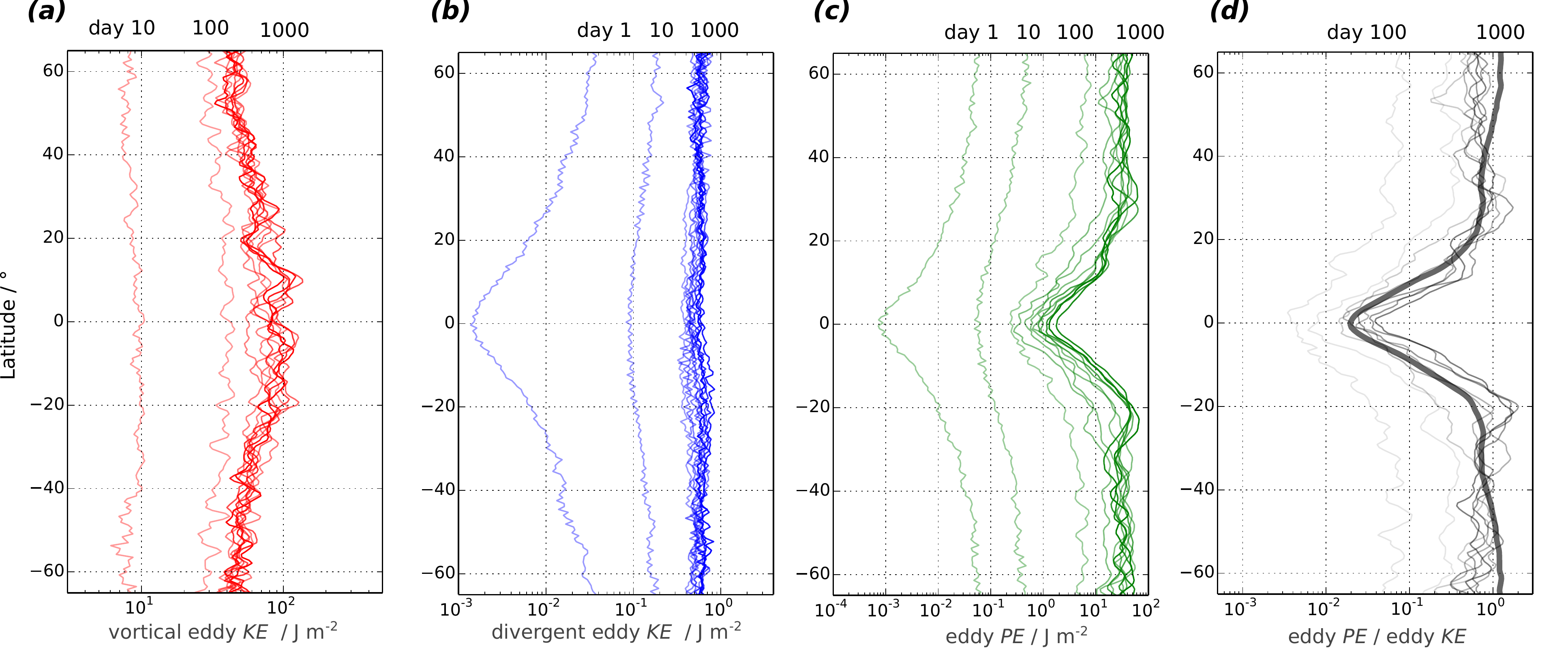}
  \caption{Zonal means of
  vortical eddy kinetic energy $E_\psi$ {\bf (a)}, divergent eddy kinetic energy $E_\phi$ {\bf (b)},
  eddy potential energy {\bf (c)},
  and a ratio of eddy potential energy to eddy kinetic energy {\bf (d)} with an ensemble average on day 1000 over 80 runs \emph{(thick gray line)}. The color scaling shows days $\in$
  $[1, 10, 100, 200, ..., 1000]$.}
  \label{fig:energy_latitudes}
\end{figure}


\noindent To describe the evolving upscale energy cascade that is excited by the stochastic forcing of vorticity at the mesoscale, we begin to characterize the latitudinal evolution of eddy energies: regarding eddies as fluctuations from a zonal mean, we denote them as $(\cdot)'$ and a zonal mean by an overbar $\overline{(\cdot)}$. Eddy kinetic and potential energy are column integrated to retrieve units of J\,m$^{-2}$ \citep{salmon1998}. 
This gives, $KE=0.5\,\rho\,H\left({u'u'}+{v'v'}\right),$
where $u'u'$ and $v'v'$ are zonal and meridional wind variance. Eddy PE is
$0.5\,\rho\,g h' h',$
with $\rho$ as a unit density of 1\,kg\,m$^{-3}$ and $h'h'$ as the height field variance.
Similarly, we define an eddy moist energy as $0.5\,\rho\,g \, L^2\, q' q'$
in units J\,m$^{-2}$ and proportional to moisture field variance\,$q'q'$. The zonal mean eddy KE in the vortical and divergent modes as well as the zonal mean eddy potential energy (PE) as functions of latitude are  shown in Figure ~\ref{fig:energy_latitudes}. In all, the vortical energy is about two orders of magnitude larger than the divergent KE. Further, vortical KE attains a maximum in the equatorial region. Apart from being locally generated, this maximum could be due to an increasing Rossby radius of deformation towards the equator \citep{theiss2004}, that allows for the inverse energy transfer to reach lower wave numbers that can hold more energy (R. Salmon, personal communication). The magnitude of the eddy PE remains nearly constant outside the tropics but falls off almost monotonically from $\pm 25^\circ$ towards the equator. As the rotational zonal mean eddy KE increases towards the equator, given that we are injecting energy into the system via vorticity forcing, this is consistent with an inefficient conversion of eddy KE to eddy PE in the tropical regions. This is captured more clearly in Figure \ref{fig:energy_latitudes}d which shows the ratio of eddy PE to rotational eddy KE. At higher latitudes this ratio nears unity but as we move into the tropics the value drops to a minimum of about $10^{-2}$ over the equator. Interestingly, the divergent kinetic energy is constant with latitude. Over time, the eddy KE composed of vortical and divergent modes reaches a quasi steady equilibrium in the fully turbulent flow. 
Compared to vortical eddy KE,  eddy PE reaches its equilibrium value a bit later (as can be seen by comparing the convergence of lines in Figure~\ref{fig:energy_latitudes}a,c), and at about half of the magnitude of 80 J m$^{-2}$ in this dry small-scale vorticity forced experiment.

\section{Moist Turbulence}

\begin{figure}
    \centering
    \includegraphics[width=1.\textwidth]{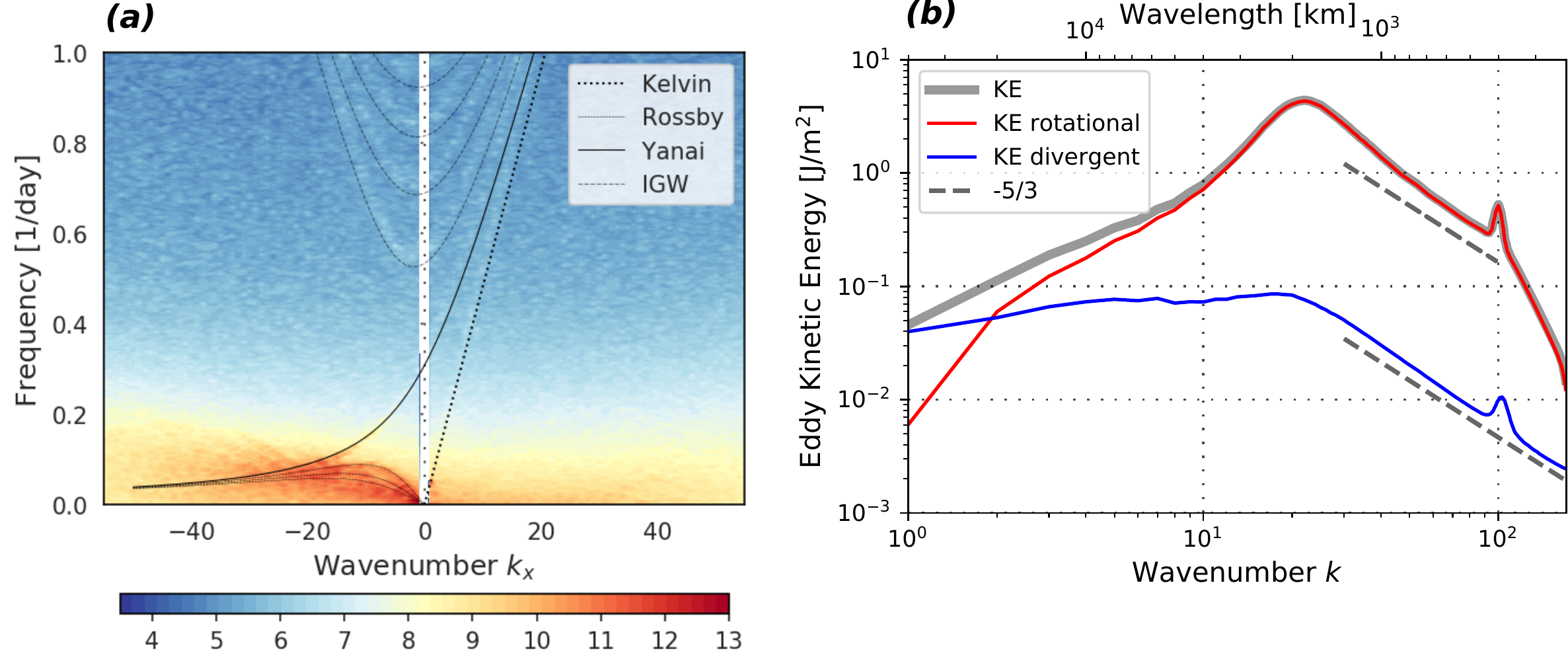}
\caption{Moist small-scale vorticity-forced: Wavenumber-frequency diagram and dispersion curves with a reduced depth of 50\,m {\bf (a)}, and eddy KE spectra with -5/3 lines for reference {\bf (c)}. }
    \label{fig:VrtFrSmallScaleMoistKeSpectra}
\end{figure}

Having established properties of the dry system in the equatorial region, we now consider the moist shallow water system with small-scale vorticity forcing. We begin with the case when the background saturation field is uniform, i.e., $q_s$ is a constant.
The corresponding wavenumber-frequency diagram is shown in Figure \ref{fig:VrtFrSmallScaleMoistKeSpectra}a. Apart from a smooth background, we see signs of heightened activity along the westward propagating Rossby wave dispersion curves, westward propagating part of the Yanai wave, and the lowest Kelvin wave modes with an appropriately reduced equivalent depth. In contrast to the dry simulation, where we saw a signature of all tropical waves (Figure \ref{fig:VrtFrSmallScaleDryKeSpectra}a), the intensity of higher Kelvin waves, eastward propagating Yanai waves, and inertia-gravity waves can hardly be differentiated from the background spectrum. Again, small-scale forcing (around 400 km), as seen in Figure~\ref{fig:VrtFrSmallScaleMoistKeSpectra}b, yields a $-5/3$ scaling with most of the energy in rotational modes. But, the inverse cascade is arrested at a smaller scale than the dry case (Figure \ref{fig:VrtFrSmallScaleDryKeSpectra}b). This is due to the fact that the equatorial deformation scale reduces in the presence of moisture by a factor of $\sqrt{\left(H-LQ\right) / H}=\sqrt{1-LQ/H}\approx 0.7$ (where $Q= \max(q_s))$, shifting the maximum of the vortical KE spectrum from wavenumber~20 of the dry case to about wavenumber~28. Once again, in the mesoscale and synoptic scales, there are signs of similar scaling in the divergent modes, albeit with a much smaller amount of energy. At the largest scales, i.e., below wavenumber 10, the vortical mode energy falls off while the divergent KE remains relatively constant. The energy of vortical and divergent KE is again of comparable magnitude over the planetary range up to wavenumber 10 --- like in spectra of the small-scale dry vorticity-forced experiment  (Figure~\ref{fig:VrtFrSmallScaleDryKeSpectra}b). As in the dry case, eddy energy in the quasi-linear vorticity-forced runs remains trapped at the forcing scale (not shown) which suggests that the upscale energy transfer is again mediated by nonlinear eddy-eddy interactions. 

\begin{figure}
  \centering
    \includegraphics[width=1.\textwidth]{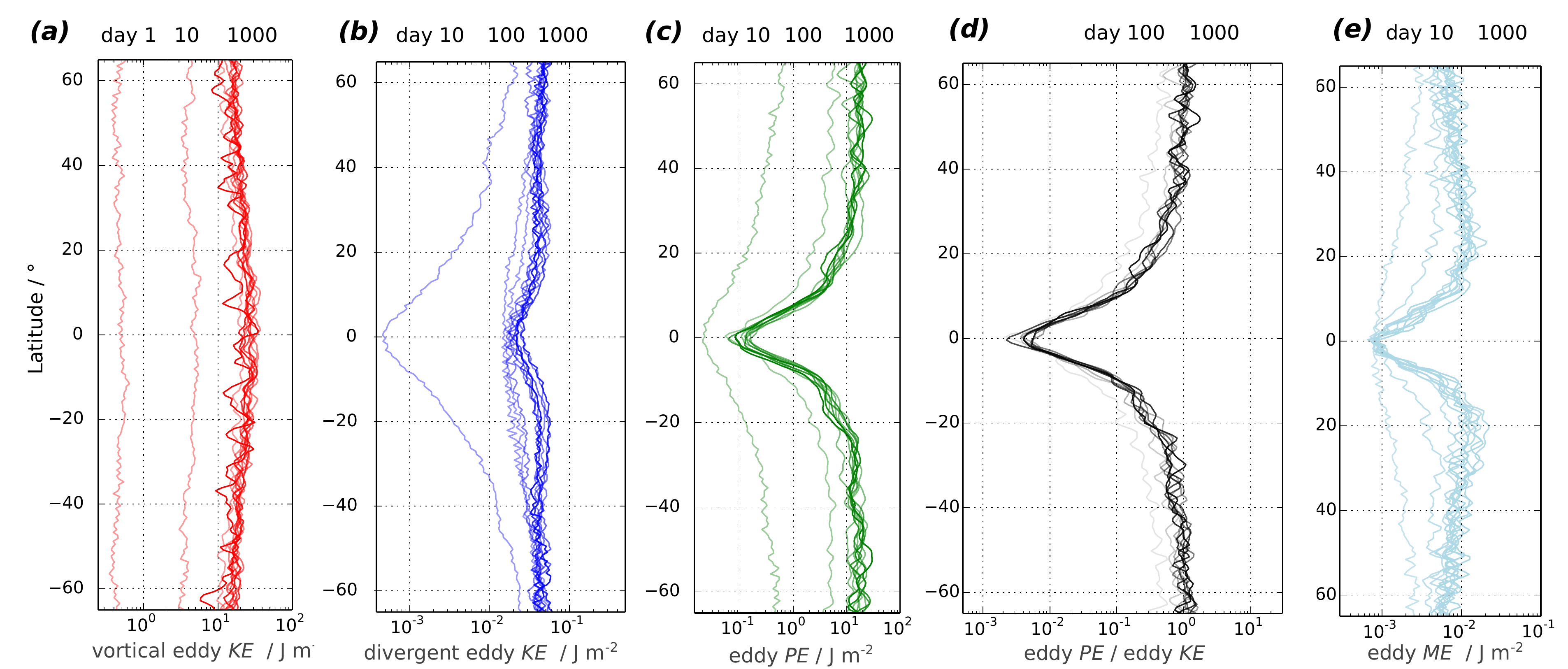}
  \caption{Similar diagnostics as Figure~\ref{fig:energy_latitudes}, but for the moist small-scale vorticity-forced run. In addition to dry diagnostics, eddy moist energy (ME) is plotted {\bf (e)}.}
  \label{fig:energy_latitudes_moist}
\end{figure}

\noindent To understand the meridional energy transport in the moist environment compared to the upscale cascade evolution in the dry run, we concentrate on the moist vorticity-forced fully non-linear run and show the vortical and divergent KE, as well as the eddy potential and moist energies as functions of latitude for different times in Figure \ref{fig:energy_latitudes_moist}. As for the dry case (Figure \ref{fig:energy_latitudes}), vortical KE dominates over the divergent part, and in fact, the difference between the two is much larger here (almost three orders of magnitude in the tropical region).
Correspondingly, the ratio between global mean vortical and divergent eddy kinetic energy is larger than in the dry runs (Table~\ref{table:experiments}).
Rotational KE increases, again, monotonically towards the tropics while PE decreases monotonically towards the equator in the tropics. The latitudinal profile of the moist energy is similar to the one of the PE, but a few orders of magnitude weaker. The main difference is seen in the steady-state divergent KE, which decreases towards the equator in the moist run, while it is constant across the tropics in the dry run. We attribute this change to weaker equatorial Kelvin and inertia-gravity waves in the moist simulation (Figure \ref{fig:VrtFrSmallScaleMoistKeSpectra}a). Furthermore, the enhanced eddy ME may contribute most to an increase in divergent eddy KE in places in the sub-tropics, where it is maximum. 

\begin{figure}
    \centering
    \includegraphics[width=1.\textwidth]{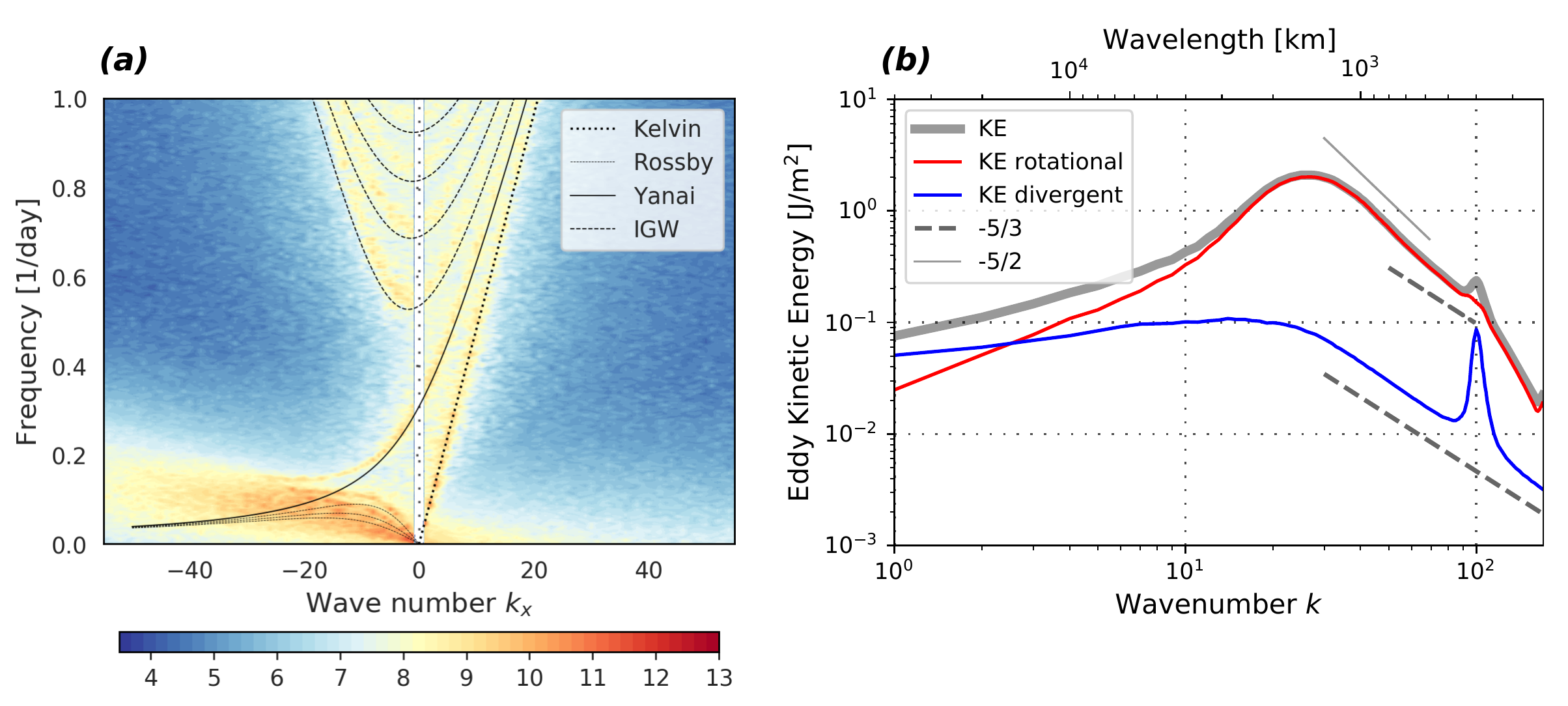}
\caption{As in Figure~\ref{fig:VrtFrSmallScaleMoistKeSpectra} but for the moist moisture-forced run, with an additional -5/2 slope line marked in {\bf (c)} for reference.}
    \label{fig:QFrSmallScaleMoistKeSpectra}
\end{figure}

\noindent When forcing the model with stochastic small-scale divergent forcing, as in the dry case, energy remains in the divergent component and is trapped at the forcing scale with no signs of interscale energy transfer or a reddish background spectrum (not shown). In contrast, stochastically forcing the moisture fields leads to vortical modes that are equally strong or orders of magnitude stronger than the divergent modes (Table~\ref{table:experiments}) in this moist environment. In fact, in addition to Rossby waves, inertia-gravity, Kelvin and Yanai waves are also produced with an appropriate reduced depth in the moisture forced runs. As a result, the wave energy is distributed over the entire family of equatorial waves (Figure~\ref{fig:QFrSmallScaleMoistKeSpectra}c). Further, the background spectrum is more pronounced than the simulated red-noise like background in the vorticity-forced runs.  The mesoscale inverse transfer appears to be influenced by moist waves or coherent structures that steepen the vortical eddy kinetic energy spectrum (Figure~\ref{fig:QFrSmallScaleMoistKeSpectra}d) to a slope of $\approx -5/2$ at comparably smaller scales than in the vorticity-forced experiment. Overall, the divergent eddy kinetic energy modes are also more energetic than in the moist vorticity-forced experiment. Interestingly, though stochastically forcing the height field shows maximum energy at the forcing scale, it also excites vortical modes that are stronger than the divergent modes. This differs from the corresponding dry runs, but is similar to results of \citet{ofer2020}, suggesting that moist dynamics introduces memory into the system, making the uncorrelated height forcing in our moist runs similar to the temporally correlated stochastic height forcing of \citet{ofer2020}. We further note that the ratio of globally averaged eddy divergent KE and PE for the divergence forcing is again 1 (Table~\ref{table:experiments}) -- as in the dry divergence forced experiment, however, in the height forced run, this ratio is 3, suggesting either that moist processes weaken the secondary excitation of divergence via height anomalies, or that the upscale cascade in this run works more efficiently on height, as well as vortical wind, relative to divergence anomalies.

\begin{figure}
  \centering
    \includegraphics[width=1.\textwidth]{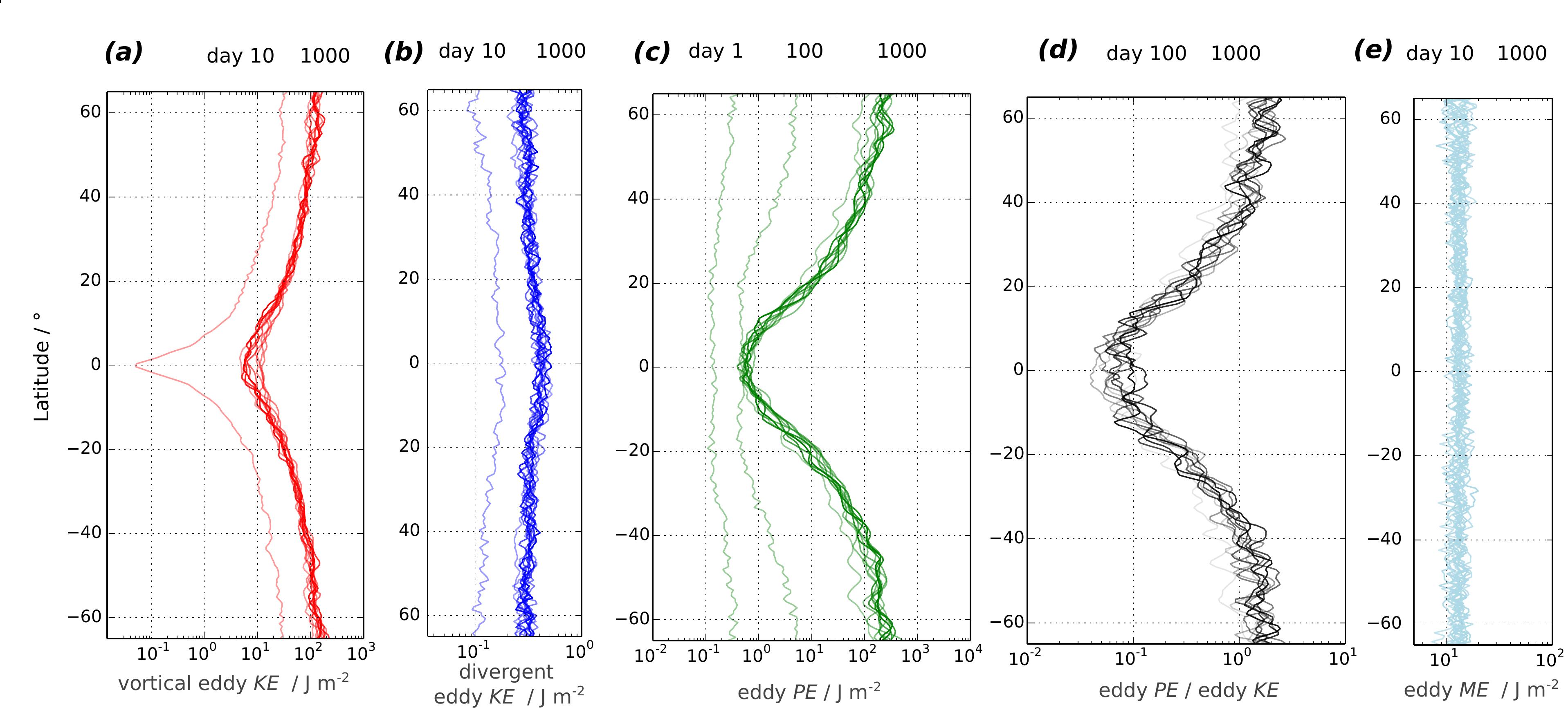}
  \caption{Similar diagnostics as Figure~\ref{fig:energy_latitudes_moist}, but for the moist small-scale moisture-forced run.}
  \label{fig:energy_latitudes_moist_fq}
\end{figure}

\noindent The latitudinal profiles of the different energy terms for the moisture-forced run are shown in Figure \ref{fig:energy_latitudes_moist_fq}. Unlike in the case of vorticity forcing, while the vortical KE is still much larger than divergent KE, it actually has a minimum, while divergent KE has a relative maximum, in the tropics. 
Since moisture only appears in the height equation, vorticity and divergence anomalies are initially excited through the height anomalies which the moisture forcing excites. Height anomalies excite divergent modes much more easily in the tropics than in midlatitudes, while rotational kinetic energy is most efficiently produced in the midlatitudes. Moreover, height anomalies are much more easily maintained in mid latitudes, where the flow is geostrophically balanced, resulting in a minimum PE in the tropics.

\subsection{Non-uniform background saturation}

\noindent Precipitation in the tropics, at a given location, is observed to be tied quite closely to the amount of column water vapor present \citep{MuEm}. Thus, given our formulation of condensation, it is natural to model the saturation field ($q_s$) as per column water vapor in the tropics. Keeping the actual distribution of precipitable water in the tropics in mind \citep{Sukhatme1}, we consider two sets of non-uniform saturation fields. The first captures the latitudinal variation in precipitable water \citep{Sukhatme2,joyQG,suhas2020}. The second takes into account both a latitudinal and longitudinal structure, and falls off in the form of a Gaussian function both meridionally and zonally from the crossing of the dateline and the equator \citep{suhas2020}. The introduction of a spatial dependence of $q_s$ is important as it introduces the possibility of condensation by means of rotational advection \citep{joyQG}, whereas for a constant saturation field condensation is only possible by means of divergence.
In fact, the influence of background moisture fields on tropical modes has been noted in shallow water studies \citep{Sobel,Sukhatme2,dias}, reanalysis data based examination of the Madden Julian Oscillation \citep{jiang} as well the birth of monsoon depressions \citep{AdamesMing}, but effects on moist turbulence have, as far as we know, not yet been studied.  



\begin{figure}
    \centering
    \includegraphics[width=1.\textwidth]{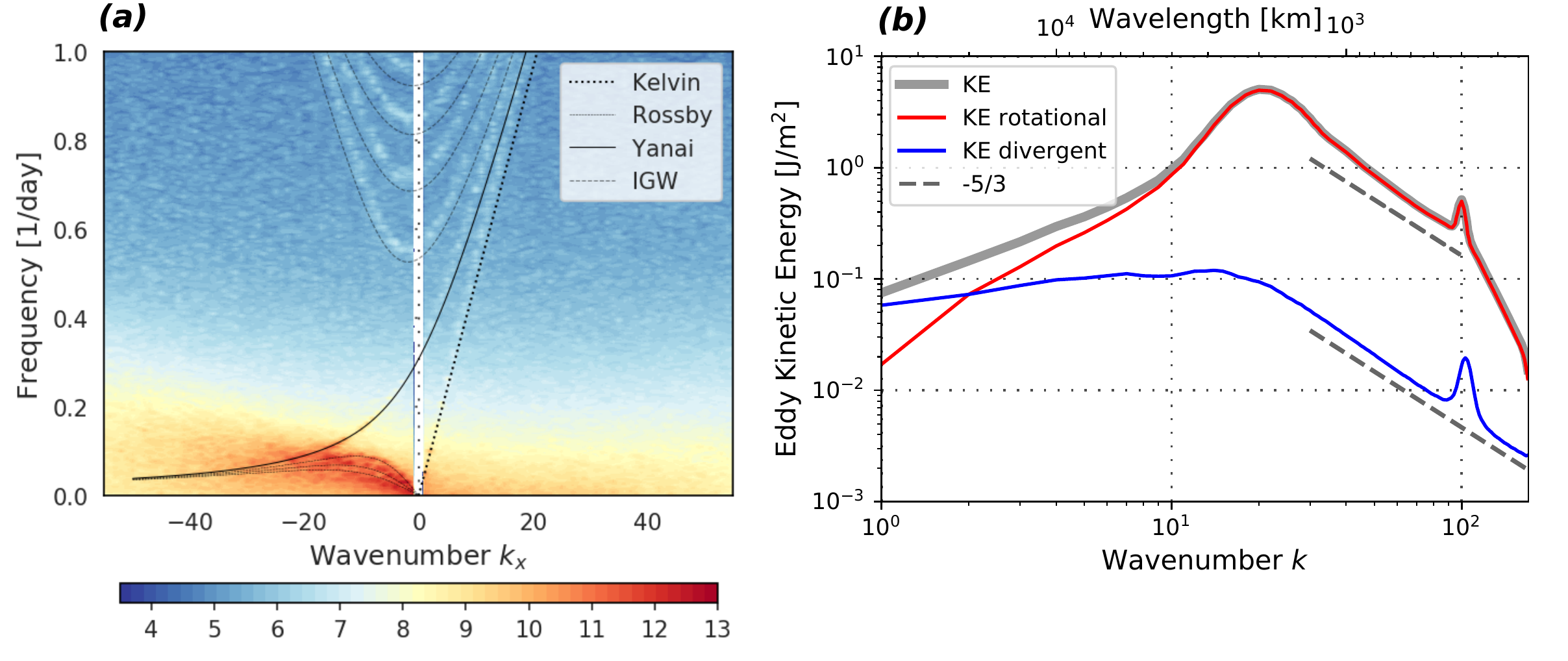}
\caption{Moist small-scale vorticity-forced run, with a stationary saturation background depending on both latitude and longitude (as described in the text): {\bf (a)} Wavenumber-frequency diagram and dispersion curves with a reduced depth of 50\,m,
    {\bf (b)} eddy KE spectra with -5/3 lines for reference.}
    \label{fig:moistlatlonsmallscalevort}
\end{figure}

\noindent We begin with $q_s$ being a function of only latitude, specifically, it has a peak at the equator and falls off as we progress poleward.
The KE spectra and wavenumber-frequency plots are almost identical to the constant $q_s$ case (hence, not shown), specifically, energy is mostly in westward propagating low-frequency modes and KE spectra are dominated by rotational modes that scale with a $-5/3$ exponent. When $q_s$ depends on both latitude and longitude\footnote{The background spectrum $q_s(lat,lon)=\exp\left(-{lat^2 \over (60^\circ)^2}  -\alpha_0 {(lon-180^\circ)^2 \over (120^\circ)^2}\right)$, where $lat$  and $lon$ are latitude and longitude in degrees, respectively. The constant $\alpha_0=0$ for the latitudinally varying, or $\alpha_0=1$ for the latitudinally and longitudinally varying background stratification.}, the wavenumber-frequency diagram and KE spectra for small-scale vorticity-forced runs are shown in Figure \ref{fig:moistlatlonsmallscalevort}. As seen in Figure \ref{fig:moistlatlonsmallscalevort}a, we again have low-frequency Rossby waves whose equivalent depth matches a rapid condensation estimate. Note that between wavenumber 10-20 there is also a signature of westward propagating mixed Rossby gravity waves. Compared to the case when the background saturation is constant (Figure~\ref{fig:VrtFrSmallScaleMoistKeSpectra}a), the excited inertia gravity waves, as well as the Kelvin and Yanai waves are more energetic. The excited higher frequency waves $>0.5\,$day$^{-1}$ are slightly faster than the linear dispersion relations for the maximum saturation value $Q$. As the saturation declines zonally, the waves experience less reduced depths > 50\,m and thus move faster. Clearly, there is a transfer of energy towards large scales  (Figure~\ref{fig:moistlatlonsmallscalevort}b) and this is arrested near the equatorial deformation radius (approximately wavenumber 30), comparable to the previous purely dry simulation. In contrast, the accumulation of divergent eddy KE at the forcing scale is less than in the purely dry case, but more than in the moist run with a constant saturation background, consistent with this case having a combination of moister and dryer regions. 
As with other cases, divergence forcing in non-uniform saturation backgrounds does not lead to interscale energy transfer (Table~\ref{table:experiments}).


\begin{figure}
    \centering
    \includegraphics[width=1.\textwidth]{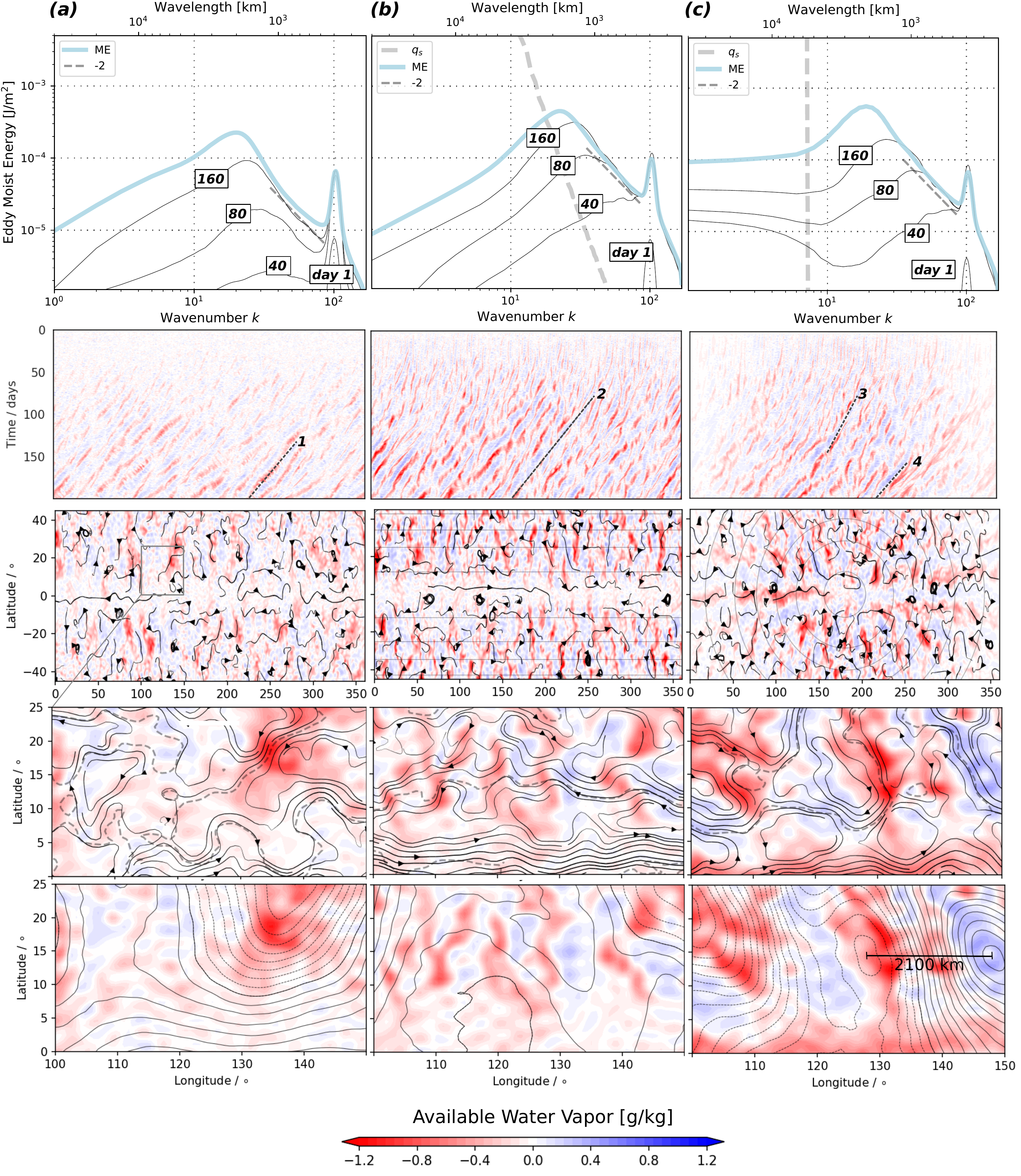}
    \caption{Spectra of eddy Moist Energy averaged from day 200-800 {\it (thick blue line)} with evolving days {\it (thin blue lines)} (row 1). All colored contours show available water vapor $q-q_s$ [g/kg] with Hovm\"oller diagrams at $20^\circ$ North evolving up to day 200 (row 2). The angular phase speed of four exemplary cases 1) -12.5$^\circ$, 2) -7.5$^\circ$, 3) -5.0$^\circ$, and 4) -8.5$^\circ$ per 10 days are plotted on top of selected wave trains {\it(dashed gray lines)}. Following below are snapshots on day 900 with streamlines of vortical wind for constant and varying saturation profiles {\it (thin gray lines)} (row 3), a zoom in of selected areas of these below with streamlines of the vortical wind with an isoline of reduced depth of 50 m {\it (dashed gray line)} (row 4), and corresponding plots with positive/negative isolines {\it (solid/dashed black lines)} of the velocity potential $\Phi$ (row 5) in:  {\bf (a)} constant, {\bf (b)} latitudinally varying,  {\bf (c)} latitudinally \& longitudinally varying saturation fields.}
    \label{fig:MsweSteadyRandomSmallScaleVrtFQsatQp}
\end{figure}

\noindent A remarkable feature of the moist runs is the aggregation of moist anomalies into coherent structures by the turbulent flow. In fact, spectra of moisture variance or eddy moist energy, shown in Figure~\ref{fig:MsweSteadyRandomSmallScaleVrtFQsatQp} (row 1), follow a power-law scaling with an approximate $-2$ exponent. 
Though note that ME is small in comparison to PE and KE$_\zeta$ which are of similar magnitude (Table \ref{table:experiments}).  Figure~\ref{fig:MsweSteadyRandomSmallScaleVrtFQsatQp} also shows the temporal evolution of the moist spectra from an initially localized form near the forcing scale to moisture variance being distributed across scales, indicating that large scale structures are being formed in the moisture field. 
The aggregation of moist anomalies is clearly visible in physical space Hovm\"oller diagrams shown in Figure~\ref{fig:MsweSteadyRandomSmallScaleVrtFQsatQp} (row 2). Aggregation takes place as time progresses, after about Day 75 in the constant $q_s$ case, and somewhat earlier in the variable $q_s$ cases. Once formed, these coherent structures systematically drift westward in the tropical region. These westward propagating systems travel about 12.5$^\circ$ for constant background saturation, over 7.5$^\circ$ with latitudinally varying background, to 5.0$^\circ$ as well as 8.5$^\circ$ for latitudinally and longitudinally varying background within 10~days. We also note that these disturbances tend to increase in speed over time, in the case where background saturation varies in both horizontal directions.
As was indicated by the KE spectra, in all cases, eddies form as energy cascades upscale. To view this mechanistically, we show snapshots of the moisture field, plotted along with different parts of the flow field (rows 3,4 and 5 of   Figure~\ref{fig:MsweSteadyRandomSmallScaleVrtFQsatQp}).
An examination of the close ups in rows 4 and 5 of Figure~\ref{fig:MsweSteadyRandomSmallScaleVrtFQsatQp} tells us that the anomalies are largest for the case when the saturation field is a function of latitude and longitude. Further, the scale of the eddies (about 2100 km; Figure~\ref{fig:MsweSteadyRandomSmallScaleVrtFQsatQp}, row 5) agrees with estimates of the moist deformation radius. 
We note a strong correlation between the dry anomaly and a local minimum of the velocity potential $\phi$ in Figure~\ref{fig:MsweSteadyRandomSmallScaleVrtFQsatQp} (row 5), which indicates a region of strong divergence as $\delta=\Delta \phi$ and proportional to $-k^2\phi$ in wave number space. Accordingly, the eddy divergent energy in varying background saturation is relatively stronger than for constant $q_s$ (Table~\ref{table:experiments}). In fact, strong divergent/convergent regions also correspond to vortical wind confluence regions, as can be deduced from Figure~\ref{fig:MsweSteadyRandomSmallScaleVrtFQsatQp} (row 4), these moist coherent structures align with regions of converging streamlines but are aided by rotational advective condensation in the cases when the saturation field depends on space. 
For example, if we examine the flow at about 130$^\circ$ longitude and 15$^\circ$\,N, the equatorward rotational flow brings parcels from a higher latitude towards the equator, resulting in a negative advective anomaly (red) as $q_s$ is higher near the equator. This anomaly enhances the convergence induced condensation/evaporation in this region as it lies between two rotational gyres. In general, from the moisture equation in (\ref{a1}), 
the evolution of anomalies ($q^+, q^- \ll q_s$) involves advection operating on $q_s$ and divergence multiplying the saturation field. 
Thus, moisture anomalies are co-located with divergence and convergence in the constant $q_s$ case, but are influenced by the vortical meridional velocity in the $q_s(lat)$ case as well as by zonal advection in the $q_s(lat,lon)$ case.

\begin{figure}
    \centering
    \includegraphics[width=.67\textwidth]{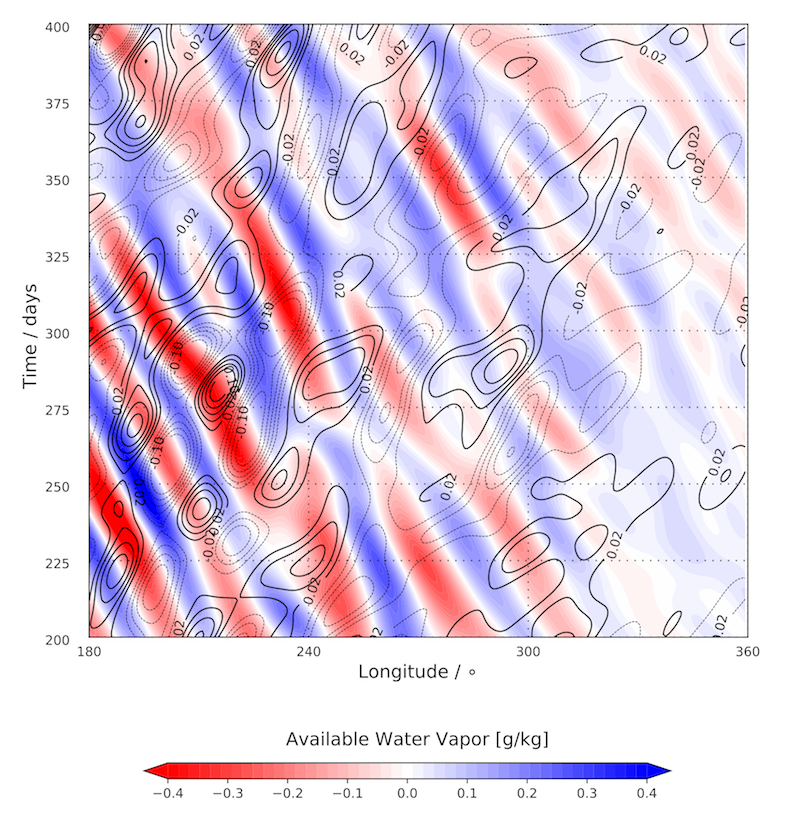}
    \caption{Moist small scale vorticity forced run with horizontally varying saturation $q_s(lat,lon)$: a Hovm\"oller diagram reveals waves in the midlatitudes at 45$^\circ$ North. Colors show westward moving moisture anomalies, and contour lines indicate negative {\it(dashed lines)} and positive {\it(solid lines)} values of eastward moving moisture anomalies from $-0.1$ to $0.1$ g/kg.}
    \label{fig:Rossby}
\end{figure}

\noindent Further towards the north, a super-position of eastward and westward propagating structures can be seen in a Hovm\"oller diagram of  moisture anomalies (Figure~\ref{fig:Rossby}). The eastward propagating moisture waves are similar in speed and longevity to the ones simulated by \citet{suhas2020} in their Figure~11, propagating 60$^\circ$ eastwards in approximately 200 days with a corresponding width of up to 30$^\circ$.
In this turbulent scenario, the westward propagating systems are relatively stronger in the midlatitudes by approximately a factor four.
Notably, these large-scale eastward propagating moisture waves only occur in the heterogeneously saturated background, where $q_s$ varies in longitude and latitude. The physical mechanism for the eastward movement is linked to the conservation of moist PV and the inversion of PV gradients in the midlatitudes as explained in \citet{suhas2020}.
Figure~\ref{fig:Rossby} further reveals that the anomalies are larger in the western part and much smaller in the eastern part of the domain - this indicates that the advection of moisture background gradients is a source of the anomalies, as the waves propagate in this direction and are altered over time. 

\section{Mean Flows}

\begin{figure}
    \centering
    \includegraphics[width=1.\textwidth]{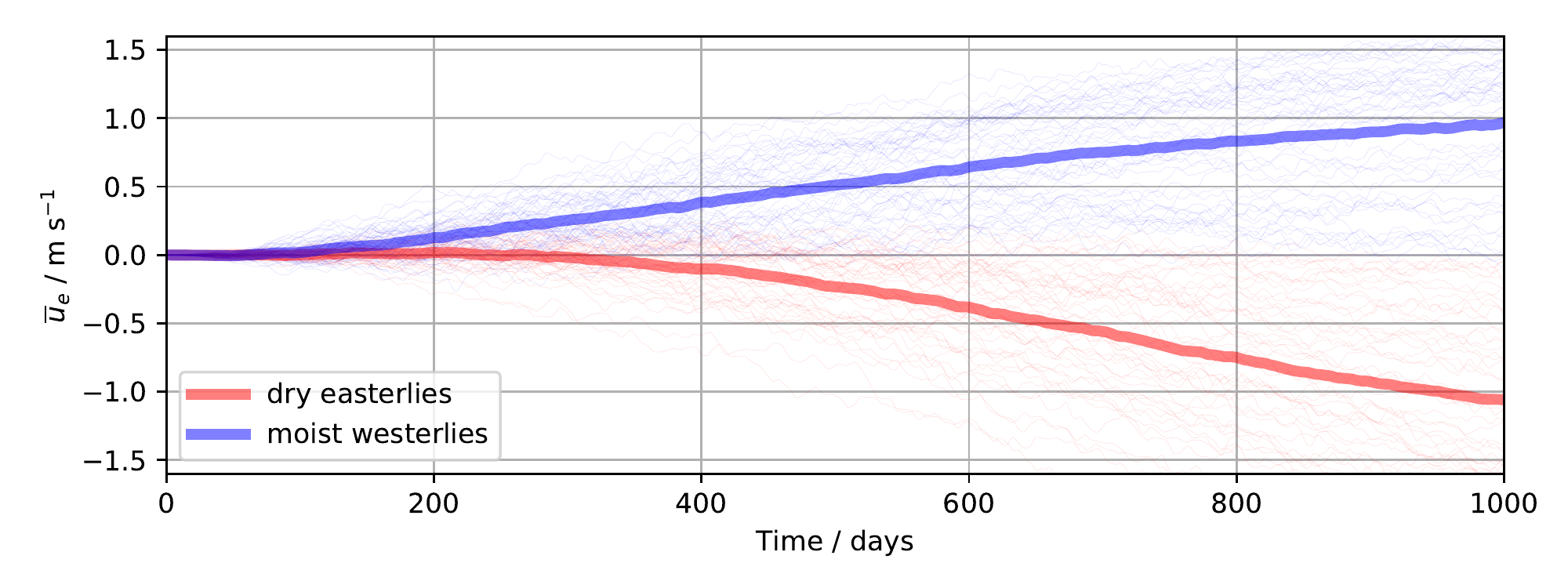}
    \caption{Time series show the zonal mean velocity at the equator $\overline{u}_e$ for the full ensemble of dry and moist simulations. The ensemble mean of easterly cases for the dry runs {\it(thick blue line)}, westerlies {\it(thick red line)} for the moist run.
    To show the ensemble spread, all individual runs are plotted in the background {\it(thin lines)}.}
    \label{fig:u_equator}
\end{figure}

\begin{figure}
    \centering
    \includegraphics[width=.75\textwidth]{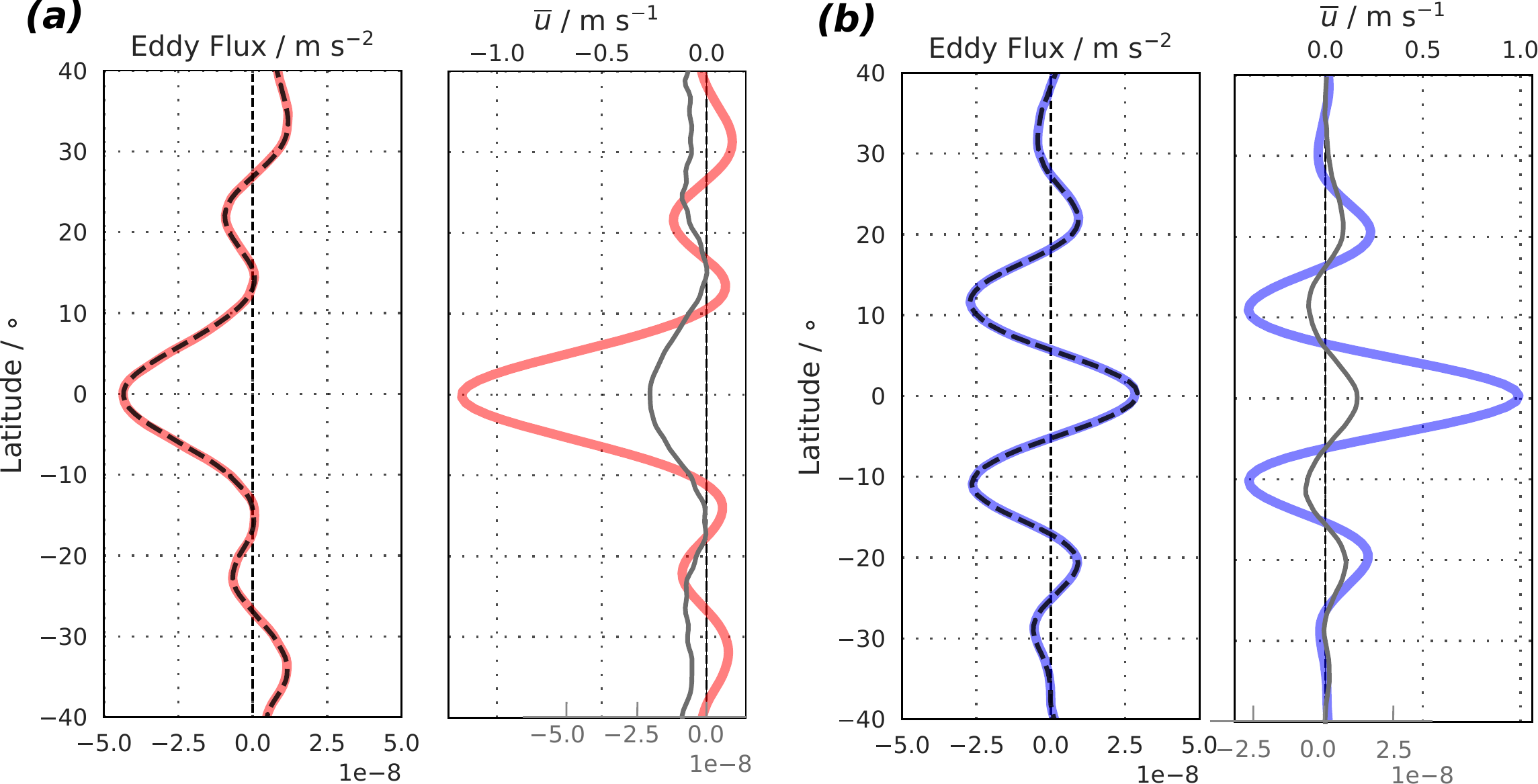}
    \caption{Eddy momentum flux and wind profile $\overline{u}$ with total momentum flux $\partial_t \overline{u}$ {\it (solid gray line)} averaged from day 800--1000. The vortical eddy flux is plotted on top {\it(thick dashed black line)}. Two cases are presented: the ensemble of dry {\bf(a)}, as well as moist {\bf(b)} vorticity-forced experiments. The colors correspond to the previous figure. }
    \label{fig:fluxes}
\end{figure}

\begin{figure}
    \centering
    \includegraphics[width=.9\textwidth]{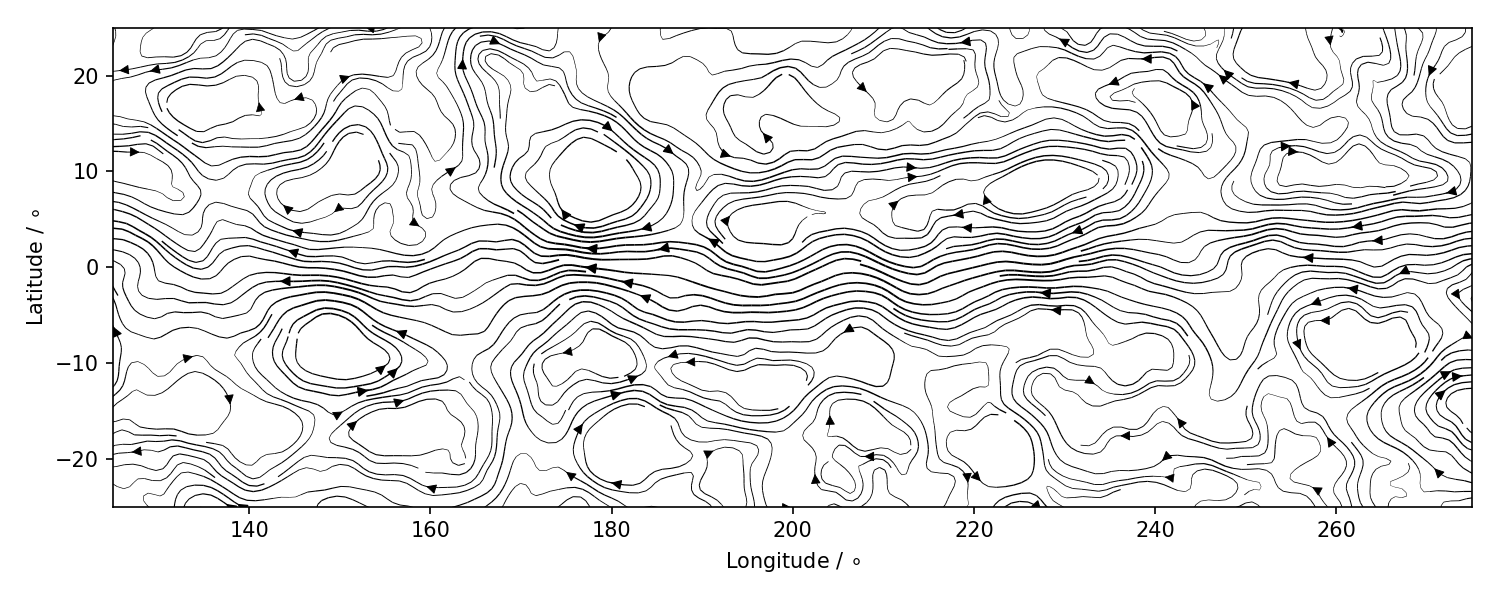}
    \caption{Dry vorticity-forced run: Streamlines reveal eddies and a zonal flow in the vortical wind ${\bf v}_\zeta$ as in Figure~\ref{fig:MsweSteadyRandomSmallScaleVrtFQsatQp}, but for the dry case between $\pm25^\circ$ North/South for a selection of 125$^\circ$ to 275$^\circ$ in longitude.}
    \label{fig:streamlines_L1}
\end{figure}

\begin{figure}
    \centering
    \includegraphics[width=.5\textwidth]{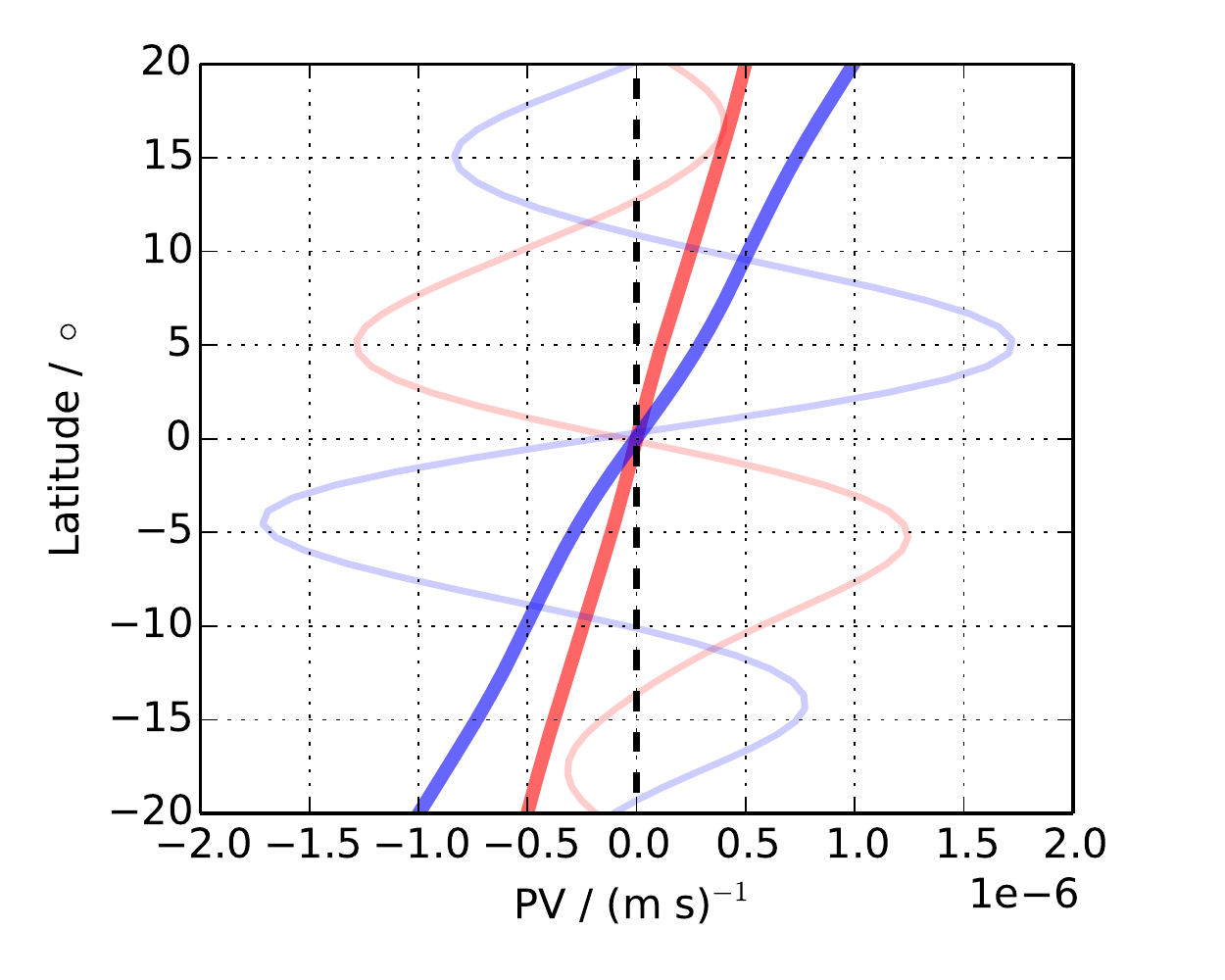}
    \caption{Ensemble means are shown as in Figure~\ref{fig:fluxes}, but for potential vorticity (PV, thick lines) and zonal mean relative vorticity ($\overline{\zeta}_r$, thin lines) for the small scale vorticity-forced runs. Here, blue and red colors are for moist and dry cases, respectively.}
    \label{fig:pv}
\end{figure}


\noindent Till now, we have focused on large scale waves and turbulence generated by the dry and moist systems with small scale forcing. In all the cases when there is an upscale energy transfer (Table~\ref{table:experiments}), we observe that the flow organizes itself into time-mean zonal equatorial jets. To examine the formation of zonal jets in more detail, we run ensembles of the dry and moist (with uniform $q_s$) vorticity-forced runs. 
Figure~\ref{fig:u_equator} shows the resulting zonal mean wind time series averaged over a band $\pm5^\circ$ centered around the equator.
As can be seen in the ensemble mean time series, there is a wide spread in the value of equatorial zonal winds, with the dry system favoring easterlies over the equator while the moist cases show a preference for equatorial super-rotation (see Table ~\ref{table:experiments} for percentages). In both cases, it takes a few hundred days for the mean zonal flow over the equator to develop and by Day 1000 the ensemble average sub- and super-rotating values are of the order of 1 m\,s$^{-1}$. The eddies have a characteristic standard deviation with same magnitude as the zonal mean flow (Table~\ref{table:experiments}). 
To understand the terms that contribute to an increase in the zonal momentum at the equator quantitatively, we perform a flux decomposition \citep{thulag,suhas2017}. 
Quite strikingly, as seen from the ensemble means, shown in Figure \ref{fig:fluxes}, in both cases, the zonal mean flow over the equator is driven by the eddy fluxes which are accelerating and decelerating in the moist and dry cases, respectively. Further, these eddy fluxes are primarily rotational in character. We note from Figure \ref{fig:u_equator} that the zonal mean zonal wind value has not reached a steady state by Day 1000 in many of the runs, and in the ensemble mean. Consistently, Figure \ref{fig:fluxes} also shows the time tendency term, calculated as the sum of all the terms of the RHS of the zonal mean zonal wind equation. We see weak but non-zero time tendencies at the equator which indeed act to strengthen the magnitude of the equatorial winds. A difference, however, between the two cases is found outside the main tropical jet- in the super-rotating runs, the time tendency has spatial structure is similar to that of the zonal wind field, so that not only is the equatorial super-rotating jet growing, but so are the easterly and westerly jets which flank it on both sides. For the dry run, on the other hand, the time tendency acts to strengthen and widen the equatorial easterly jet, and shift or weaken the secondary jets flanking winds. This is consistent with the secondary jets being stronger in the moist run ensemble mean. Indeed, the ensemble spread of zonal wind in the moist runs is much smaller than in the dry runs (not shown). 

\noindent The positive equatorial eddy-momentum flux divergence in the super-rotating runs indicates a tropical wave source. Since we are forcing waves at all latitudes, this tropical wave source suggests the turbulent excitation of tropical relative to extra-tropical vortical waves is more efficient in the moist runs. 
This is consistent with the dominance of equatorial Rossby waves seen in the wavenumber-frequency diagram (Figure~\ref{fig:VrtFrSmallScaleMoistKeSpectra}a) as well as the tilt of the mainly cyclonic gyres in the physical space picture in rows 4 and 5 of Figure~\ref{fig:MsweSteadyRandomSmallScaleVrtFQsatQp}. An examination of Hovm\"oller diagrams of the different ensemble members suggests that many of the strong super-rotating jets converge from wider to narrower confined jets over time (not shown).  
We further note that the easterly jets which form to the north and south (slightly equatorward of $\pm 20^\circ$) are co-located with the local maxima of divergent eddy kinetic energy (Figure~\ref{fig:energy_latitudes_moist}b). The decelerating nature of the eddy flux in the dry runs bears further scrutiny. As can be seen in Figure~\ref{fig:streamlines_L1}, the streamlines of the vortical wind expand over a wider region and a negative momentum flux is established in the dry equatorial region leading a weak easterly wind. Further, gyres within $\pm 10^\circ$ North/South in the dry case are mostly anticyclonic in nature. Interestingly, the cyclonic and anticyclonic nature of the rotational eddies, in the moist and dry runs, results in a tropical circulation which tries to homogenize PV in the dry case, but exaggerates PV gradients in the moist simulations (Figure~\ref{fig:pv}).

\begin{figure}
    \centering
    \includegraphics[width=.85\textwidth]{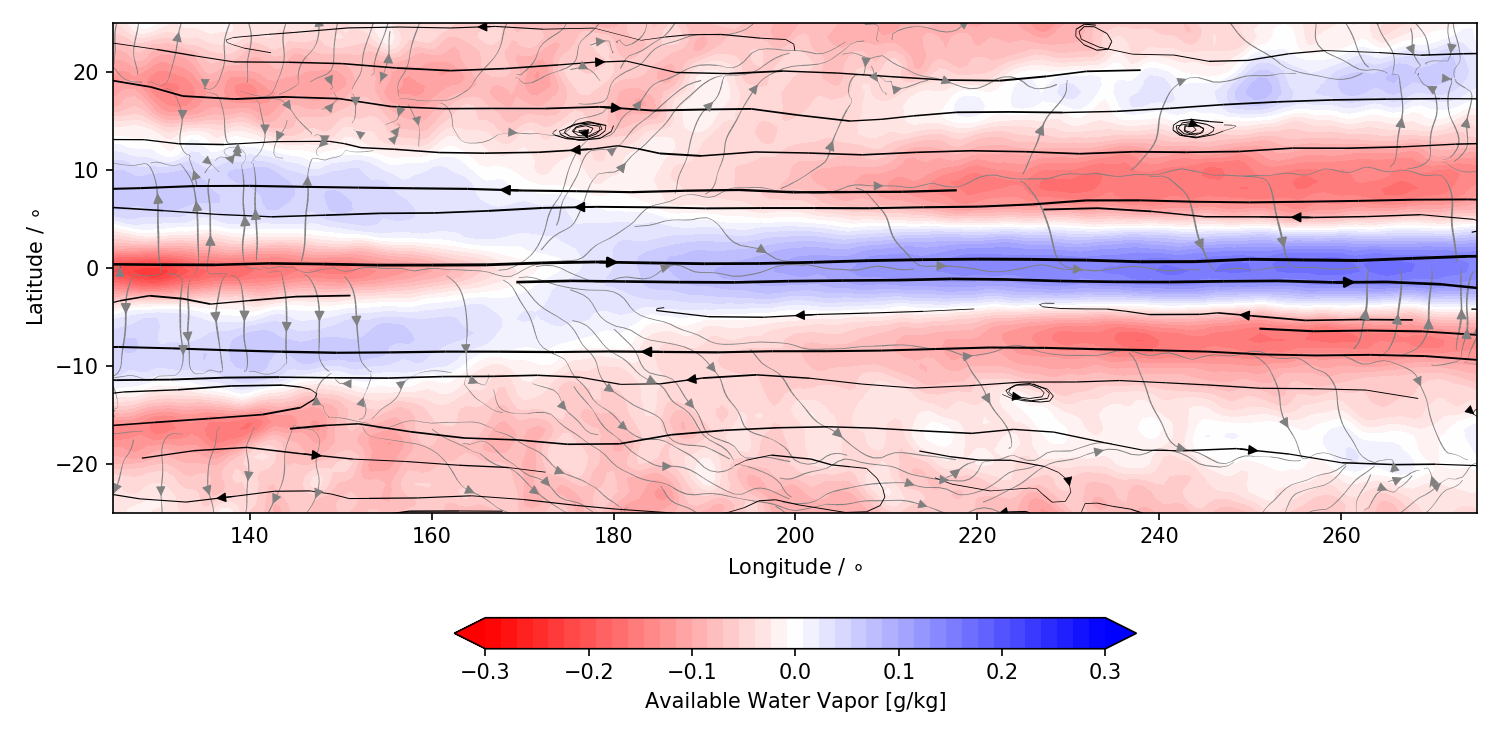}
    \caption{Moist vorticity-forced run with $q_s(lat,lon)$: A large-scale pattern across the equator exists in the time mean of the moisture anomaly $q'$ {\it (plotted is the deviation from the tropical mean)}.
    Streamlines of time mean divergent {\it (thin gray lines)} and vortical wind {\it (thick black lines)} are overlaid.}
    \label{fig:qslatlon_timemean}
\end{figure}

\noindent In the run with a saturation field that depends on latitude and longitude (i.e., $q_s(lat,lon)$), 
the presence of a zonally varying mean moisture field alongside a zonal mean flow induces a stationary planetary scale moisture wave in the equatorial region. As can be see in  Figure~\ref{fig:qslatlon_timemean}, the stationary wave divergence and the zonal advection along the equator in the eastern and western halves lead to the negative (red) and positive (blue) moisture anomalies, respectively, with moisture damping balancing both. Note that moisture anomalies are similarly forced more poleward, around $\pm 10^o$ latitude, in the region of the secondary easterly jets (though note that divergence does not contribute to the forcing of the moist anomalies westward of $180^o$, resulting in a slightly weaker moisture anomaly there).
The fundamental source of this stationary tropical moist wave is the heterogeneous background saturation field, whereas the mean flow that is needed to excite and sustain the wave is purely driven by the stochastically forced upscale energy cascade from the mesoscale. Overall, this strong connection between divergent and vortical wind with the moisture field emphasizes their inextricable nature and ranges from the mesoscale forcing to planetary scale waves. 

\begin{table}
\begin{center}
\begin{tabular}{l | c | c | c | c | c | c | c | c | c  }
    Forcing    & moist & $k^{-5/3}$ & $\langle KE_\zeta \rangle \over \langle KE_\delta\rangle$ & $\langle PE\rangle \over \langle KE_\delta\rangle$ & $\langle ME\rangle \over \langle KE_\delta\rangle$ & runs & $\overline{u}_e$ &  $\sigma_u / \overline{u}_e$ & $\overline{\delta}_e$ \\
                         \hline
    $f_\zeta$~ $\square$    & ~ & $\times$ & 100 & 80 & ~ & 160 & < 0 for 68\,\%  & 1/2 & > 0 \\ 
    $f_\delta$   & ~ & ~ & 1/50 & 1 ~ & & 1 & 0 &  $\infty$ & 0 \\ 
    $f_h$        & ~ & ~ & 1/10 & 1 ~ & & 1 & 0 &  $\infty$ & 0 \\
    $f_\zeta$, quasi-linear    & ~ & ~ & 1000 & 30 & ~ & 2 & 0 &  $\infty$ & 0   \\
                         \hline
    $f_\zeta$ $\bigcirc~\square$    & $\times$ & $\times$ & 400 & 100 & 1/10 & 80 & > 0 for 72\,\% & 1/2 & > 0    \\
    $f_\delta$   & $\times$ & ~ & 1/30 & 1 & 1/20 & 1 & 0  & $\infty$ & 0  \\
    $f_h$   & $\times$ & ~ & 3 & 3 & 1/20 & 2 & < 0 & 1 & 0  \\
    $f_q$   & $\times$ & $\times$ & 200 & 100 & 20 & 80 & < 0 & 1 & > 0 \\
    $f_\zeta$, quasi-linear    & $\times$ & ~ & 1000 & 100 & 1/20 & 2 & 0 &  $\infty$ & 0   \\
                         \hline
    $f_\zeta$, varying $q_s$ & $\times$ & $\times$ & 300 & 100 & 1/80 & 2 & > 0   & 1/2 & > 0  \\
    $f_\delta$, varying $q_s$ & $\times$ & ~ & 1/20 & 1 & 1/100 & 2 & 0 & $\infty$ & 0 
  \end{tabular}
\caption{Overview of ensemble runs to indicate which experiments produce a $k^{-5/3}$ upscale cascade, their
associated zonal mean wind at the equator ($\overline{u}_e$), standard deviation $\sigma_u$ and zonal mean divergence of the meridional wind at the equator $\overline{\delta}_e=\langle \partial_y \overline{v}\rangle_e$ averaged from day 800-1000. Global mean (denoted by $\langle \cdot \rangle$) ratios of vortical eddy KE $\langle KE_\zeta\rangle$, eddy PE $\langle PE\rangle$, and eddy ME $\langle ME\rangle$ to divergent eddy KE $\langle KE_\delta\rangle$. Two height forcing experiments $f_h$ were conducted in the moist environment with varying forcing amplitudes, two vorticity-forcing ones varying saturation field $q_s$ as a function of latitude only, as well as of latitude and longitude. 
Moreover,
one experiment was undertaken linearizing the dry shallow water equations with respect to a constant
zonal mean wind, and another one with respect to a latitudinally varying zonal mean wind - each for dry and moist environments. 
Varying strengths of the diabatic heating were considered for the vorticity-forced nonlinear moist runs $\bigcirc$ increasing the moist coupling from 25\,\% to 100\,\%, for 80 ensemble members in each latent heat release scenario. The square $\square$ further indicates that one of these ensemble members ran for at least 10,000 days, while all the other runs were conducted for 1000 days.}
\label{table:experiments}
\end{center}
\end{table}

\section{Summary \& Outlook}

A detailed numerical investigation is performed using the spherical dry and moist shallow water systems to simultaneously investigate equatorial turbulence and large-scale tropical waves. We successfully simulated the co-existence of a turbulent upscale kinetic energy cascade with a $-5/3$ slope and equatorial waves in the tropical region. In particular, both moist and dry runs with small scale vorticity forcing (at approximately 400 km) result in a rotational mode dominated inverse transfer regime that persists up to the respective deformation scale. The rotational scaling is similar to classic geostrophic turbulence \citep{Charney}, and its dominance suggests that this inverse transfer involves equatorial Rossby triads \citep{ripa}. Even though the divergent component has much lesser energy than the rotational part at synoptic scales, it too scales as a power-law with a $-5/3$ exponent. Mechanism denial quasi-linear simulations do not show an upscale transfer of energy thus highlighting the dynamical importance of nonlinear eddy-eddy interactions. At planetary scales the divergent modes have comparable energy to the rotational modes and wavenumber-dispersion diagrams show the footprint of tropical waves. While the entire family of waves is seen in the dry runs, only low frequency modes stand out from the background spectrum in the moist cases. 
While previous studies have simulated the emergence of equatorial waves through random forcing \cite{ofer2020}, or a non-linear convective feedback \cite{Yang}, this work shows the presence of clearly pronounced and discrete equatorial waves in co-existence with a fully developed turbulent flow.

\noindent Notably, the existence of an inverse transfer of energy in the vortical component crucially depends on the type of variable that is being forced continuously with random white noise. While uncorrelated divergent forcing leads to an accumulation of energy at the forcing scale, vorticity and moisture forcing enhance an efficient non-linear transport of energy from the mesoscale to the planetary scale. The main difference in the moist and dry cases is seen under forcing of the height field. In the dry case, this behaves like divergence forcing with no upscale transfer and an accumulation of energy at the forcing scale. Whereas, a -5/3 cascade is excited in vortical modes, when height is forced in the moist system and reaches up to the amplitude of the energy that accumulates at the forcing scale in the height forced run (not shown). When stochastically forcing moisture, vortical modes are excited and they follow a $-5/3$ with a steeper (close to $-5/2$ exponent) near the deformation scale. In a sense, given the similarity of our moist height and moisture forced runs with the temporally correlated height forcing in the dry system \cite{ofer2020}, it appears that moisture lends some memory to the system and allows for a transfer of energy to the rotational component which then flows to larger scales.

\noindent The upscale energy transfer of KE in moist runs is accompanied by self-aggregation of moisture in the tropics. Specifically, with small scale vorticity forcing, it takes about a 100 days for coherent moist structures to appear. Further, once formed, the aggregates propagate westward in the tropics with speeds of the order of a few meters per second. With regard to their energy, moisture variance scales with a slope of $-2$ throughout the mesoscale. Moreover, moisture anomalies are largest in the case when the background saturation field depends on both longitude and latitude. Here, these anomalies are supported by advection and convergence, and are situated between the rotational gyres that dominate the tropics. It is important to note that coherent moist anomalies appear spontaneously without radiative feedbacks or surface fluxes that are sometimes deemed to be essential ingredients in aggregation \citep[for example,][]{wingreview}. In the case when the saturation field depends on latitude and longitude, along with low-frequency tropical waves \& turbulence, as expected from moist PV conservation \citep{suhas2020}, an eastward moving, large-scale moisture wave emerges in the midlatitudes. 



\noindent In addition to waves and turbulence, a systematic zonal mean zonal flow develops over the equator. This flow takes about 500 days to develop and is easterly and westerly in the dry and moist cases, respectively. Interestingly, easterly flows at the equator coincide with relatively strong active Kelvin waves, while westerly flows exhibit strong Rossby wave activity. A momentum budget shows that the eddy flux drives the zonal mean zonal flow in both moist and dry runs, moreover this flux is primarily rotational in character and the tilt of the eddies is consistent with the sense of the momentum flux in the two scenarios. In all, rotational eddies in the moist and dry runs are cyclonic and anticyclonic in character, respectively, thus the tropical circulation generated tends to homogenize PV in the dry case, but exaggerates PV gradients in the moist simulations. This mostly rotational zonal mean flow results in a large-scale, tropical stationary moisture wave in the case when the saturation field depends on latitude and longitude. The spatially heterogeneous nature of the saturation is essential for this stationary moist wave, and so is the time mean zonal flow that arises from an inverse energy transfer among the rotational eddies. Taken together, not only do these experiments showcase the concurrence of large scale equatorial waves and fully developed turbulence in the tropical region, they highlight the the interdependence of the rotational and divergent portions of the flow with the dynamically interactive moisture field.

\noindent We have focused on small-scale forcing in this work, and it would be interesting to simulate energy transfer in a moist model that can resolve baroclinic instability and thus, provides a canonical source for an enstrophy forward cascade. 
Not only will this allow us to probe possible transitions from a $-3$ to $-5/3$ scaling in KE, the presence of large-scale baroclinic waves can moreover be a source for low frequency equatorial waves \cite{wedi2010}. Generally, in a multiple layer model, studying the vertical circulation in space-time spectra of equatorial waves as done for numerical weather prediction reanalysis data (George Kiladis, personal communication) could be of interest, when vertical transport of moisture is explicitly resolved in dynamically coupled layers.

\section*{Acknowledgements}

This work was conducted in a joint research project between the Israeli Science Foundation (grant number 2713/17) and University Grants Commission, India (grant number F6-3/2018). J. Schr\"ottle wants to thank Adam Sobel and Rick Salmon for insightful discussions in the spring of 2019, as well as for the questions of all participants of the Meeting 'Physics at the equator: from the lab to the stars' in Lyon in October 2019. 

\clearpage

\section*{Appendix A) Isotropy within the inertial subrange}

\noindent Two-dimensional spectra within the inertial subrange of shallow water height $h$, exhibit an isotropy in $x-$ and $y-$direction.
Spectra of kinetic energy are slightly elongated in $x-$direction. Spectra of the stochastic forcing are isotropic and show a maximum at
the forcing scale. At the forcing scale of wave number 100 on the sphere, random modes are selected within the range of total wave number 
$k \in [98, 102]$. Decomposed in Fourier modes with random amplitudes, the forcing $\bf f$ can be written as follows:

$${\bf f} = (f_\zeta, f_\delta, f_h, f_q) = (\hat{a}_\zeta, \hat{a}_\delta, \hat{a}_h, \hat{a}_q)\, \exp(i 2 \pi A),$$
where $A$ is a complex two dimensional matrix over all spherical wave numbers with entries at wave numbers corresponding to
total wave number $k$ in the forcing range and zero outside. Within the forcing range, entries are chosen randomly as complex numbers of
magnitude~1. In this way, the wave magnitude between $[-\hat{a},\hat{a}]$ and phase between [0$^\circ$,\,360$^\circ$] vary randomly.
Amplitudes $\hat{a}=\hat{a}_\zeta$, $\hat{a}_\delta$, $\hat{a}_h$, or $\hat{a}_q$ are positive real numbers.

\begin{figure}
  \centering
    \includegraphics[width=.85\textwidth]{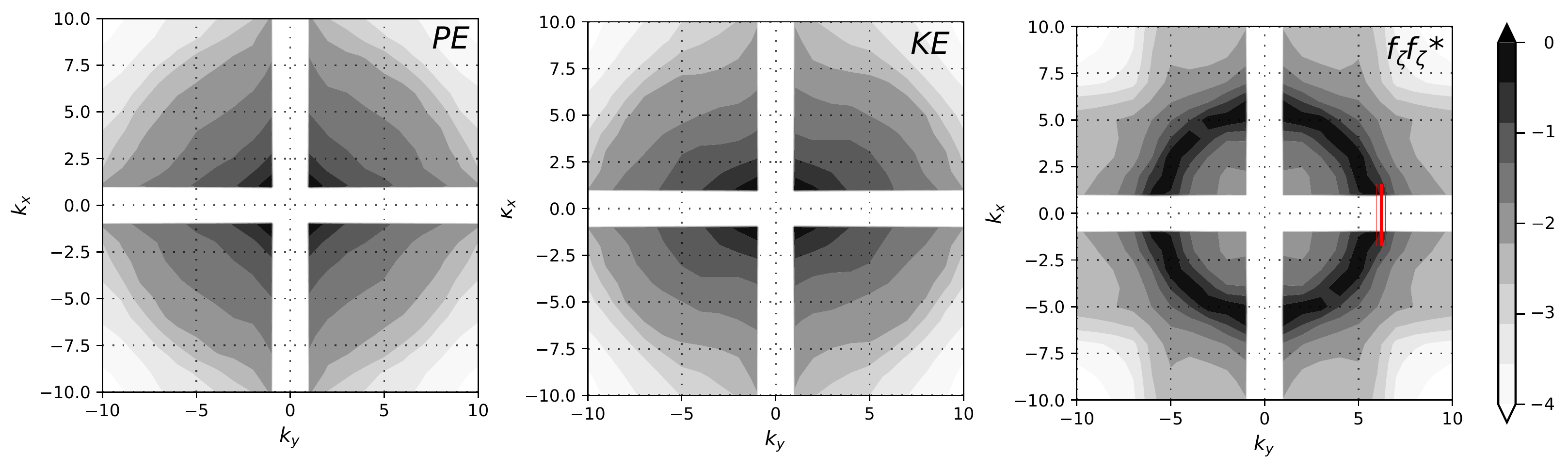}
  \caption{Horizontal spectra of potential energy $PE$ {\bf (left)}, kinetic energy $KE$ {\bf (middle)}, and small-scale vorticity forcing with its complex conjugate $f_\zeta^{\,}f_\zeta^*$ {\bf (right)} all normalized by their respective field maximum. The scale of stochastic forcing is indicated {\it (red lines)}.}
  \label{fig:spectra_kx_ky}
\end{figure}

\section*{Appendix~B) Intermittency due to coherent structures}

\begin{figure}
  \centering
    \includegraphics[width=.32\textwidth]{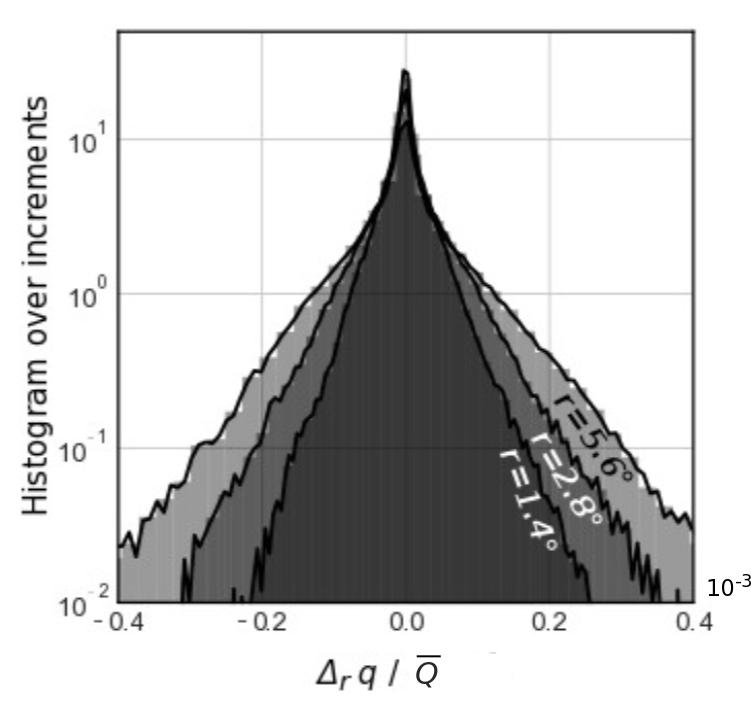}
  \caption{Distributions of $\Delta_r q$ in the passive tracer limit ($L=0$).}
  \label{fig:histo}
\end{figure}

\noindent At small scale, the flow is intermittent as can be seen in distribution of scalar increments (Figure~\ref{fig:histo}).
The intermittency is due to gradients at the scale $r$ of the $increments$. Interestingly, this distributions resembles
first-guess departure distributions in satellite data-assimilation that are strongly non-Gaussian.
In contrast, probability distributions of all scalar variables $u$, $v$, $q$, and $h$ in the shallow water simulations
resemble Gaussian distributions as expected form the central limit theorem for randomly forced fully developed two-dimensional turbulence \citep{kuksin2006}.

\section*{Appendix~C) Partition of energy into rotational and divergent modes}

\noindent The sum of vortical kinetic energy $KE_\zeta$ and divergent kinetic energy $KE_\delta$ gives the total  kinetic energy $KE$ in spectral space. This can be deduced from Helmholtz decomposition of the velocity field in two-dimensional Fourier space. To retrieve correct units, the kinetic energy is normalized by domain mean height $\overline{H}$ and unit density $\rho=1\,$kg/m$^3$,
\begin{align*}
KE/\rho\overline{H} = {1 \over 2} \left( u^2 + v^2\right) = {1 \over 2} \left(\left(u_\zeta + u_\delta\right)^2 + \left(v_\zeta + v_\delta\right)^2\right)=~~~~\\
= {1 \over 2} \left(u_\zeta^2 + 2\,u_\zeta u_\delta + u_\delta^2 + v_\zeta^2 + 2\,v_\zeta v_\delta + v_\delta^2\right)=~~~~~~~\\
= {1 \over 2} \left(u_\zeta^2 + v_\zeta^2\right)  + {1 \over 2} \left(u_\delta^2 + v_\delta^2\right) + u_\zeta u_\delta + v_\zeta v_\delta.~~~~~~~~~
\end{align*}
Using a Helmholtz decomposition with streamfunction $\psi$ for vortical modes and potential $\phi$ for divergent modes, this equation can be rewritten. Thereby, $(u_\zeta, v_\zeta)$ are the vortical modes of the wind and $(u_\delta, v_\delta)$ are its divergent modes,
\begin{align*}
KE/\rho\overline{H} = {1 \over 2} \nabla \psi \cdot \nabla \psi + {1 \over 2} \nabla \phi \cdot \nabla \phi - \partial_y \psi  \partial_x \phi  +  \partial_x \psi \partial_y \phi.~~~~~~~~~
\end{align*}
In Fourier Space the terms encapsulating vortical modes and divergent modes sum to zero,
\begin{align*}
\widetilde{KE} = {1 \over 2} \left(k_x^2 + k_y^2\right)\tilde{\psi}^2 + {1 \over 2} \left(k_x^2 + k_y^2\right)\tilde{\phi}^2 \underbrace{- k_y k_x \tilde{\psi} \tilde{\phi} + k_x k_y \tilde{\psi} \tilde{\phi}}_{=0}= \widetilde{KE}_\zeta + \widetilde{KE}_\delta,
\end{align*}
where $\tilde{\psi}$ and $\tilde{\phi}$ are the amplitudes of streamfunction and velocity potential in Fourier Space. Similarly,  $\widetilde{KE}$, $\widetilde{KE}_{\zeta}$, $\widetilde{KE}_{\delta}$ are the amplitudes of kinetic energy, vortical kinetic energy, and divergent kinetic energy. In consequence, the total kinetic energy is a direct sum of vortical and divergent eddy kinetic energy as diagnosed previously in the spectral decomposition.

\newpage
\bibliographystyle{apalike}
\bibliography{clouds.bib}

\end{document}